\documentclass{article}
\usepackage{iclr2027_conference,times}

\usepackage{amsmath,amsfonts,bm}

\def\eqref#1{equation~\ref{#1}}

\def\1{\bm{1}}

\DeclareMathAlphabet{\mathsfit}{\encodingdefault}{\sfdefault}{m}{sl}
\SetMathAlphabet{\mathsfit}{bold}{\encodingdefault}{\sfdefault}{bx}{n}

\usepackage[hypertexnames=false]{hyperref}
\usepackage{url}
\usepackage{amsmath}
\usepackage{amssymb}
\usepackage{graphicx}
\usepackage{array}
\usepackage{tabularx}
\usepackage{enumitem}
\usepackage{multirow}
\usepackage{booktabs}
\usepackage{subcaption}
\usepackage[table]{xcolor}
\usepackage{float}
\usepackage{tikz}
\usetikzlibrary{arrows.meta,positioning,shapes.geometric,calc}

\definecolor{bestfill}{HTML}{F4CCCC}
\definecolor{secondfill}{HTML}{FCE5CD}
\newcommand{\best}[1]{\cellcolor{bestfill}\textbf{#1}}
\newcommand{\second}[1]{\cellcolor{secondfill}\textbf{#1}}
\newcommand{\seclink}[2]{\hyperref[#1]{\S#2}}

\title{\raggedright
Beyond the Transcriptome: Chromatin-\\
Informed Prediction of Cell-State-Dependent\\
Perturbation Responses}

\iclrfinalcopy
\author{\normalfont
\textbf{Jiafa Ruan}\textsuperscript{1,2,3}
\quad
\textbf{Chenyang He}\textsuperscript{1}
\quad
\textbf{Ruijie Quan}\textsuperscript{1,2,3}
\\[0.6ex]
\textbf{Fanglei Xue}\textsuperscript{4}
\quad
\textbf{Zongxin Yang}\textsuperscript{2}
\quad
\textbf{Yi Yang}\textsuperscript{1,2,3}
\\[1.0ex]
\textsuperscript{1}Institute of Fundamental and Transdisciplinary Research,
Zhejiang University\\
\textsuperscript{2}College of Artificial Intelligence,
Zhejiang University\\
\textsuperscript{3}State Key Laboratory of Brain-Machine Intelligence,
Zhejiang University\\
\textsuperscript{4}University of Washington\\}

\begin{document}
\maketitle
\lhead{Preprint}

\begin{abstract}
Predicting transcriptional responses to genetic perturbations is central to
understanding gene function.
Existing predictors primarily rely on transcriptomic measurements, although
chromatin accessibility provides complementary information about the cellular
context in which perturbations act.
Using this information requires linking chromatin context to specific
perturbations and accounting for baseline differences between independently
sampled control and perturbed populations.
We propose \textbf{ChromaPert}, a chromatin-informed framework for predicting
state-dependent perturbation responses.
ChromaPert combines molecular and DNA-informed locus priors with paired
control RNA--ATAC features to jointly represent target identity and measured
cellular context.
Its Chromatin-Guided Response Router (CGRR) builds a source bank of
control-derived expression offsets and retrieves relevant sources using
perturbation/locus or control-ATAC similarity.
These offsets account for baseline differences, while conditional
transport flow learns the remaining response.
Across three unseen-target and held-out-state settings on K562 CAT-ATAC and
Perturb-Multiome, ChromaPert achieves the highest all-gene response correlation
among evaluated methods. When transferring known perturbations to held-out
states, it improves correlation
by 23.2\% across all genes and 46.2\% for the twenty most responsive genes over
the strongest respective baselines.
Correct RNA--ATAC pairing improves recovery of chromatin-associated response
slopes over shuffled pairing, while cross-state predictions retain
state-specific responses to the same perturbation.
\end{abstract}

\section{Introduction}

Predicting transcriptional responses to perturbations is central to
functional genomics and gene-function characterization.
Single-cell CRISPR and combinatorial screens provide large-scale
measurements~\citep{dixit2016perturb,norman2019exploring}, but their coverage
remains incomplete across perturbations and cellular contexts. This gap
motivates \emph{in silico} prediction for unseen perturbations and cellular
backgrounds.

Most perturbation predictors represent cellular state using transcriptomic
profiles from single-cell RNA sequencing (scRNA-seq), which measures RNA
abundance as a readout of gene expression. They learn expression shifts,
exploit gene relationships, or model mappings from control to perturbed
cells~\citep{lotfollahi2019scgen,
lotfollahi2023predicting,roohani2024predicting,bunne2023learning,
klein2025cellflow}.
Yet transcriptomic state alone may miss regulatory context encoded by
\textbf{chromatin accessibility}---the physical accessibility of DNA
packaged with proteins. Single-cell ATAC sequencing (scATAC-seq) measures
this complementary signal~\citep{buenrostro2015single}, which may improve
perturbation prediction in two ways. First, accessibility at response-
relevant regulatory regions can influence transcription-factor recruitment
and downstream transcription; for example, it shapes glucocorticoid-receptor
binding after hormone stimulation~\citep{john2011chromatin,
biddie2011transcription}. Second, it can distinguish cellular backgrounds
with similar RNA profiles but different regulatory states, as observed in
paired RNA--ATAC measurements of mouse skin, where basal-cell populations
showed distinct accessibility patterns~\citep{hao2021integrated}.
Together, these findings motivate chromatin-informed modeling of perturbation
responses.

Using chromatin for perturbation prediction raises two
challenges.
\textbf{i) How can the RNA--chromatin background be linked to the response
to a specific perturbation?}
Control RNA--ATAC measurements describe gene expression and chromatin
accessibility in unperturbed cells, whereas the prediction target is the
expression change associated with a specific perturbation.
The model must connect the RNA--chromatin background
with the perturbation and learn how they jointly shape the transcriptional
response.
\textbf{However, because single-cell profiling is typically destructive, this
relationship must be learned from separately sampled control and perturbed
populations} rather than longitudinal before--after measurements of the same
cells~\citep{bunne2023learning}.
This introduces the second challenge:
\textbf{ii) How can pre-existing expression differences be distinguished
from perturbation-associated changes in these unpaired populations?}
Control and perturbed populations may have different expression baselines.
When reference controls do not represent the appropriate baseline,
particularly when state-specific perturbed expression is compared with
controls pooled across states, the observed difference can contain both
pre-existing baseline shifts and perturbation-associated changes.
The model must therefore account for these control-derived baseline
differences rather than learning them as perturbation responses.

To address these challenges, we introduce \textbf{ChromaPert}, a
chromatin-informed framework for predicting perturbation expression
(Figure~\ref{fig:cgrflow_overview}). To link cellular background to the specific
perturbation, its \textbf{chromatin-informed perturbation condition} combines
molecular and DNA-informed locus priors with paired control RNA--ATAC features.
Protein and RNA-token embeddings encode target identity, while promoter-sequence
and regulatory annotations describe its locus context. These complementary
priors relate unseen targets to training perturbations; control RNA--ATAC
features specify the cellular background in which they act.
To account for baseline differences, its
\textbf{Chromatin-Guided Response Router (CGRR)} builds a source bank of
control-derived expression offsets. It routes queries by perturbation/locus
similarity for unseen-target prediction or control-ATAC similarity for
cross-state transfer. This decomposition separates baseline variation from
the response learned by conditional transport flow, which reconstructs
post-perturbation expression relative to matched controls for each perturbation
in its corresponding cellular background.

ChromaPert achieves the highest all-gene response correlation among evaluated
methods in unseen-target prediction on K562
CAT-ATAC~\citep{shevade2025simultaneous} and
Perturb-Multiome~\citep{martin2025transcription}, and in cross-state transfer on
the latter.
Relative held-out-state gains reach \textbf{23.2\%} across all genes and
\textbf{46.2\%} for the twenty most responsive genes over the strongest
respective baselines (Table~\ref{tab:gse_cross_state_comparison}).
Ablations support the model components; RNA--ATAC pairing and cross-state
analyses show preserved chromatin-associated and state-dependent response
structure (Table~\ref{tab:ablation_set}; Figure~\ref{fig:atac_dependence}).

\par\noindent
Our main contributions are summarized as follows:
\begin{itemize}[leftmargin=*,itemsep=0.05em,topsep=0pt,parsep=0pt,partopsep=0pt]

\item \textbf{Chromatin-informed perturbation conditioning.}
We integrate molecular and DNA-informed locus priors with paired control
RNA--ATAC features, allowing the perturbation condition to adapt to the
measured RNA--chromatin context.

\item \textbf{Chromatin-guided response routing.}
CGRR builds response sources from control-derived expression offsets and
performs query-dependent routing, separating pre-existing baseline variation
from the response subsequently modeled by conditional transport flow.

\item \textbf{Predictive and response-structure validation.}
ChromaPert improves perturbation-response prediction across unseen-perturbation
and cross-state settings, while controlled ATAC and response-structure analyses
support the predictive value of chromatin-informed conditioning.

\end{itemize}

\section{Related Work}

\paragraph{Single-cell perturbation prediction.}
Latent-shift, compositional, and transport models learn intervention effects
from independent control and perturbed populations
~\citep{lotfollahi2019scgen,lotfollahi2023predicting,bunne2023learning,
yu2025perturbnet,adduri2026predicting}. Gene relations, regulatory networks,
causal structure, and sparse mechanism shifts provide biological constraints
~\citep{roohani2024predicting,kamimoto2023dissecting,wu2022predicting,
bereket2023modelling,chi2025grape,aibar2017scenic}. Foundation models supply
pretrained gene or cell representations
~\citep{theodoris2023transfer,cui2024scgpt,hao2024large,
yang2024genecompass,zeng2025cellfm,wang2026illuminating}. ChromaPert uses target
representations and control chromatin for context-dependent source estimation
and response prediction across perturbation targets and cellular backgrounds.

\paragraph{Generative modeling of unpaired responses.}
Generative models, including diffusion models, predict responses to unseen interventions
~\citep{qi2024predicting,chi2025unlasting,yuan2026perturbdiff,ruan2026beyond}.
CellFlow uses flow matching, and scDFM adds distribution-level alignment
~\citep{klein2025cellflow,yu2026scdfm}. Conditional flow matching
parameterizes transport between distributions
~\citep{lipman2022flow}, but its learning target depends on the
response definition and control reference. ChromaPert defines responses relative
to a pooled training-control reference, subtracts a source component derived
from control-state offsets, and fits a train-only conditional transport flow to
the remaining response. Its source-on candidate combines the routed source
component with the decoded transport-flow displacement before reliability
gating and matched-control reconstruction.

\paragraph{Chromatin-aware and multimodal modeling.}
Chromatin accessibility complements RNA-derived cell state
~\citep{buenrostro2015single,ma2020chromatin}.
ArchR, weighted-neighbor analysis, and CLM-X integrate regulatory or multimodal
information~\citep{granja2021archr,hao2021integrated,li2026clm}; Perturb-Multiome
and CAT-ATAC provide paired perturbation profiles
~\mbox{\citep{martin2025transcription,shevade2025simultaneous}}.
CellOracle uses multimodal regulatory networks to simulate transcription-factor
perturbations~\citep{kamimoto2023dissecting}. ChromaPert uses target/locus or
control-ATAC similarity, according to the evaluation setting, to route each
query to expression offsets derived from training control states. These
control-derived offsets, not known perturbation effects, form the source prior for
conditional transport-flow prediction.

\section{Method}
\label{sec:method}
\label{sec:model_overview}

\begin{figure}[t]
    \centering
    \includegraphics[width=\linewidth]{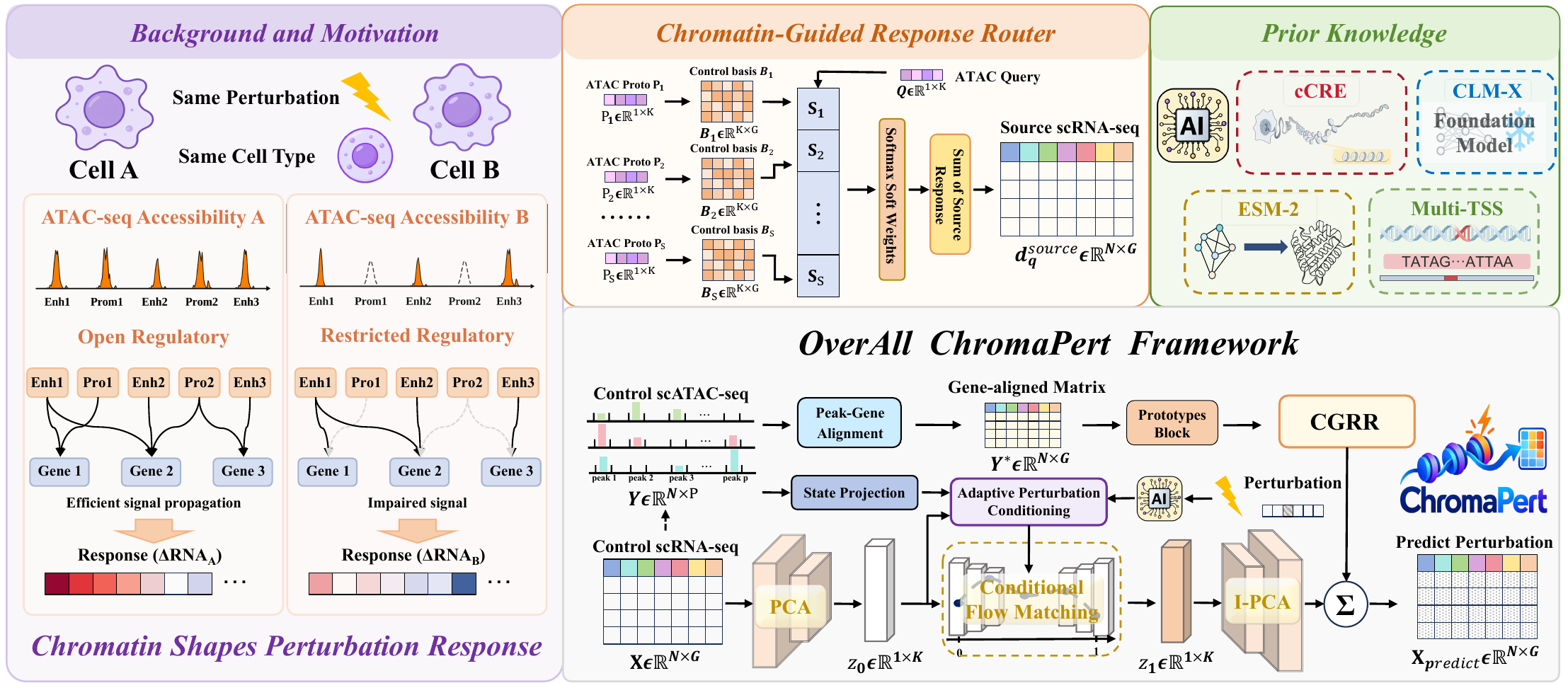}
    \caption{\textbf{ChromaPert overview.} \seclink{desc:model_overview}{3} \textbf{Left:} chromatin context
    shapes the response to the same perturbation. \textbf{Top:} CGRR pairs
    control-RNA offsets $B_s$ with ATAC prototypes $P_s$, using target/locus
    retrieval for Target-LOO and ATAC routing for Cross-State.
    \textbf{Bottom:} target/locus priors condition the flow, with paired
    control RNA--ATAC summaries added to the state-aware branch.
    \textbf{Right:} predicted responses are
    reconstructed relative to the matched-control RNA baseline.}
    \label{fig:cgrflow_overview}
\end{figure}

\noindent\textbf{Overview.}\label{desc:model_overview} As shown in Figure~\ref{fig:cgrflow_overview},
ChromaPert predicts post-perturbation expression by connecting
perturbation identity with measured RNA--chromatin context. It comprises
three components: (i) \textbf{chromatin-informed perturbation conditioning}
combines molecular and locus priors with control RNA--ATAC summaries;
(ii) \textbf{chromatin-guided response routing} represents control-derived
baseline variation and selects relevant training sources; and
(iii) \textbf{conditional transport flow} learns the remaining response and
reconstructs post-perturbation expression relative to matched controls.

\noindent\textbf{Prediction setup.} Let $p$ denote a perturbation target and
$q$ a query cellular context. Control and perturbed cells are independently
sampled populations, without before--after cell pairing. For training
condition $i$ with perturbed mean $\mu_i^{\mathrm{pert}}$, we define the
$G$-gene response relative to the mean training-control expression
$\bar b_{\mathrm{tr}}$:
\begin{equation}
    \Delta_i=\mu_i^{\mathrm{pert}}-\bar b_{\mathrm{tr}}.
    \label{eq:pooled_delta}
\end{equation}
At inference, matched-control RNA $b_q$, control-ATAC context $Q_q$, target
$p$, and static target/locus resources $\Pi_p$ determine a response
$\widehat\Delta_{p,q}$ and post-perturbation expression
$\widehat\mu^{\mathrm{pert}}_{p,q}=b_q+\widehat\Delta_{p,q}$.
Target-LOO holds out $p$; Cross-State holds out state $q$ while retaining $p$
in other training states, with query controls available only at prediction.

\subsection{Chromatin-Informed Perturbation Conditioning}
\label{sec:state_encoding}

\noindent\textbf{Target and locus priors.} For target $p$, frozen
64-dimensional ESM-2 protein~\citep{lin2023evolutionary} and CLM-X
RNA-token~\citep{li2026clm} embeddings form $u_p$. A 96-dimensional
sequence/locus block and a 128-dimensional regulatory projection form
$\ell_p$:
\begin{equation}
    \begin{aligned}
    u_p&=[e_p^{\mathrm{ESM2}}\Vert e_p^{\mathrm{CLMX}}]\in\mathbb R^{128},
    &\ell_p&=[f_p^{\mathrm{seq}}\Vert f_p^{\mathrm{reg}}]\in\mathbb R^{224},\\
    x_p&=[u_p\Vert\ell_p]\in\mathbb R^{352}.
    \end{aligned}
    \label{eq:perturb_condition}
\end{equation}
The sequence block uses TSS-centered features from GENCODE/RefTSS on
GRCh38~\citep{frankish2021gencode,abugessaisa2019reftss,schneider2017evaluation};
the regulatory block combines guide-library descriptors, ENCODE cCREs, and
JASPAR motifs~\citep{martin2025transcription,encode2020expanded,castro2022jaspar}.
These priors are fixed across cells for a given target.

\noindent\textbf{Control context and condition assembly.} Paired control
RNA--ATAC measurements~\citep{buenrostro2015single,hao2021integrated} provide
the context $h_s=[h_s^{\mathrm{RNA}}\Vert h_s^{\mathrm{ATAC}}]\in\mathbb R^{160}$
for state or control group $s$: 96 RNA dimensions and 64 chromatin dimensions,
the latter obtained by control-only TF--IDF/LSI~\citep{stuart2021single}.
The router and pooled flow append the explicit locus block to $x_p$;
the state-aware branch additionally receives $h_s$:
\begin{equation}
    c_p^{\mathrm{pool}}=[x_p\Vert\ell_p]\in\mathbb R^{576},
    \qquad
    c_{p,s}^{\mathrm{state}}=[x_p\Vert\ell_p\Vert h_s]\in\mathbb R^{736}.
    \label{eq:condition_dimensions}
\end{equation}
Thus, the state-aware condition adapts the same target priors to the measured
RNA--chromatin environment. All blocks use training-fold standardization;
missing priors are zero-filled.

\subsection{Chromatin-Guided Response Routing}
\label{sec:cgrr}

\noindent\textbf{Control-derived source bank.} Pooled responses can mix
perturbation effects with pre-existing control-state differences. CGRR
represents the latter through control-RNA offsets:
\begin{equation}
    B_s=b_s-\bar b_{\mathrm{tr}}\in\mathbb R^G,
    \label{eq:source_basis}
\end{equation}
where $b_s$ is the control mean of training group $s$. These offsets are
fixed during flow training and contain no observed perturbation effects.
Let $\mathcal P_{\mathrm{tr}}$ and $\mathcal S_{\mathrm{tr}}$ denote the
training targets and source groups.

\noindent\textbf{Target-based routing.} For Target-LOO, CGRR retrieves
training perturbations by target/locus similarity and combines their observed
state compositions $m_i(s)$:
\begin{equation}
    w_i^{\mathrm{feat}}(p)=\operatorname{softmax}_i
    \!\left(-\frac{d(c_p^{\mathrm{pool}},c_i^{\mathrm{pool}})}
    {\tau_{\mathrm{feat}}}\right),
    \qquad
    m_p^{\mathrm{feat}}(s)=\sum_{i\in\mathcal P_{\mathrm{tr}}}
    w_i^{\mathrm{feat}}(p)m_i(s),
    \label{eq:cgrr_feature}
\end{equation}
where $d$ is Euclidean distance after training-fold standardization and
$\tau_{\mathrm{feat}}=0.20$.

\noindent\textbf{Chromatin-based routing.} For Cross-State, clean-control
ATAC $Q_q$ routes the query among training-state ATAC prototypes $P_s$:
\begin{equation}
    m_q^{\mathrm{ATAC}}(s)=\operatorname{softmax}_s
    \!\left(\frac{\operatorname{cos}(Q_q,P_s)}{\tau_{\mathrm{ATAC}}}\right),
    \qquad \tau_{\mathrm{ATAC}}=0.15.
    \label{eq:cgrr_cross_state}
\end{equation}

\noindent\textbf{Baseline decomposition.} The routed source offset is
\begin{equation}
    d_{p,q}^{\mathrm{source}}=
    \sum_{s\in\mathcal S_{\mathrm{tr}}}m_q(s)B_s,
    \label{eq:cgrr_source}
\end{equation}
with $m_q=m_p^{\mathrm{feat}}$ for Target-LOO and
$m_q=m_q^{\mathrm{ATAC}}$ for Cross-State. Training uses observed compositions
$m_i(s)$, with control frequencies as the fallback for insufficient support,
to define the source-adjusted response:
\begin{equation}
    \widetilde\Delta_i=\Delta_i-d_i^{\mathrm{source}}.
    \label{eq:source_adjusted_target}
\end{equation}
This removes the composition-weighted control offset; for a single-state
condition, $\widetilde\Delta_i=\mu_i^{\mathrm{pert}}-b_s$.
Cross-State reuses the ATAC weights to transfer state-centered responses,
without source addback in the final readout.

\subsection{Conditional Transport and Response Reconstruction}
\label{sec:delta_flow}
\label{sec:composition}

\noindent\textbf{Pooled response branch.} A conditional flow learns
$\widetilde\Delta_i$. Its decoded prediction $a_p$, blended with the training
mean using $\rho=0.18$, gives two Target-LOO candidates:
\begin{equation}
    \Delta_{p,q}^{\mathrm{on}}=
    \mathcal C(d_{p,q}^{\mathrm{source}}+a_p),
    \qquad \Delta_{p,q}^{\mathrm{off}}=a_p.
    \label{eq:transport_base}
\end{equation}
Source addback restores the pooled-response convention. The stabilization
operator $\mathcal C$ preserves supported genes and blends unsupported genes
toward the pooled training response with $\alpha=0.10$
(Equation~\ref{eq:candidate_stabilization}). A separate flow trained directly
on $\Delta_i$ supplies a stabilized clean-flow candidate.

\noindent\textbf{State-centered response branch.} When state structure is
available, a separate flow uses $c_{p,s}^{\mathrm{state}}$ and matched-control
response targets:
\begin{equation}
    r_{i,s}=\mu_{i,s}^{\mathrm{pert}}-b_s,
    \qquad
    A_{p,s}=(1-\rho)a_{p,s}^{\mathrm{state}}+\rho\bar r_s,
    \label{eq:state_response}
\end{equation}
where $a_{p,s}^{\mathrm{state}}$ is the decoded displacement and $\bar r_s$
the training mean response for state $s$. Rows with fewer than 20 perturbed
cells fall back to $\Delta_i$. Aggregating $A_{p,s}$ yields the
\emph{statebase}, which uses no source addback.

\noindent\textbf{Conditional flow matching.} All branches share a fixed
PCA map fitted on training controls. For a branch target
$y_j\in\{\widetilde\Delta_i,r_{i,s},\Delta_i\}$, the latent endpoints are
\begin{equation}
    z_0=E_{\mathrm{PCA}}(\bar b_{\mathrm{tr}}),
    \qquad
    z_{1,j}=E_{\mathrm{PCA}}(\bar b_{\mathrm{tr}}+y_j).
    \label{eq:flow_endpoints}
\end{equation}
With $t\sim\mathcal U(0,1)$, conditional flow
matching~\citep{lipman2022flow} uses
\begin{equation}
    \begin{aligned}
    z_{t,j}&=(1-t)z_0+t z_{1,j},
    \qquad u_j^{*}=z_{1,j}-z_0,\\
    \mathcal L_{\mathrm{CFM}}&=\mathbb E_{j,t}
    \!\left[\left\|v_\phi(z_{t,j},t\mid c_j)-u_j^{*}\right\|_2^2\right].
    \end{aligned}
    \label{eq:delta_path}
\end{equation}
Training uses fold-derived reliability weights. At inference, eight Euler steps
from $z_0$ give the displacement
$D_{\mathrm{PCA}}(z^{(8)})-D_{\mathrm{PCA}}(z_0)$, canceling the affine PCA
offset.

\noindent\textbf{Protocol-specific reconstruction.} For Target-LOO, a
train-only reliability gate combines statebase, source-on, source-off, and
clean-flow candidates into $\Delta_{p,q}^{\mathrm{gate}}$. It uses performance
on nearby training targets, agreement with a training-response estimate, and
source-shift norm guards. Cross-State bypasses the gate and transfers
state-centered responses through ATAC routing:
\begin{equation}
    \widehat\mu^{\mathrm{pert}}_{p,q}=b_q+
    \begin{cases}
    \Delta_{p,q}^{\mathrm{gate}}, & \text{Target-LOO},\\[0.3em]
    \displaystyle\sum_{s\in\mathcal S_{\mathrm{tr}}}
    m_q^{\mathrm{ATAC}}(s)A_{p,s}, & \text{Cross-State}.
    \end{cases}
    \label{eq:response_composition}
\end{equation}
The Cross-State branch adds the held-out state's matched-control mean $b_q$
once, with no separate routed-source term.

\noindent\textbf{Population readout.} For Target-LOO, a common gate-induced
adjustment preserves differences among state-specific predictions. Supplied
state proportions $\pi_q(s)$ then give
\begin{equation}
    \widehat\mu^{\mathrm{pert}}_{p,q}=
    \sum_s\pi_q(s)\bigl[b_{q,s}+\widehat\Delta_{p,q,s}\bigr],
    \qquad \sum_s\pi_q(s)=1.
    \label{eq:population_readout}
\end{equation}
State labels and counts supply these proportions; perturbed expression is
used only for evaluation. All model fitting, PCA, standardization, and source
bank construction use the outer-training split. Query controls provide the
RNA baseline and, for Cross-State, the ATAC route. This gives the predicted
post-perturbation expression; further details appear in
Appendices~\ref{app:priors}--\ref{app:implementation}.

\begin{table}[t]
\caption{Single-cell-line unseen-target perturbation prediction on K562.
Metrics are averaged over the 18 held-out targets. \seclink{desc:k562}{4.2}}
\label{tab:k562_comparison}
\centering
\scriptsize
\setlength{\tabcolsep}{2.2pt}
\renewcommand{\arraystretch}{1.08}
\resizebox{\textwidth}{!}{%
\begin{tabular}{cccccccccc}
\toprule
\multirow[c]{2}{*}{\textbf{Family}} & \multirow[c]{2}{*}{\textbf{Method}} & \multirow[c]{2}{*}{\textbf{Public.}} &
\multicolumn{2}{c}{\textbf{Mean-Response Shape}} & \multicolumn{3}{c}{\textbf{Pointwise Accuracy}} & \multicolumn{2}{c}{\textbf{Transport / Direction}} \\
\cmidrule(lr){4-5}\cmidrule(lr){6-8}\cmidrule(lr){9-10}
\textbf{} & \textbf{} & \textbf{} &
\textbf{Pearson$_{\Delta}$ $\uparrow$} &
\textbf{Pearson$_{\Delta,\mathrm{DE}}$ $\uparrow$} &
\textbf{$L2$ $\downarrow$} & \textbf{MSE $\downarrow$} &
\textbf{MAE $\downarrow$} & \textbf{W1 $\downarrow$} &
\textbf{Disc. Cos. $\uparrow$} \\
\midrule
Statistical & \textit{Control} & -- & -0.0001 & -0.0346 & 1.6135 & 0.0013 & 0.0194 & 6.9185 & 0.0001 \\
\midrule
 & GEARS & \textit{Nat. Biotechnol.'24} & 0.0878 & 0.1308 & 2.4023 & 0.0029 & 0.0371 & 6.9013 & 0.0825 \\
Graph & CellOracle & \textit{Nature'23} & 0.0572 & 0.0872 & 1.9800 & 0.0020 & 0.0245 & 7.6820 & 0.0525 \\
 & GraphVCI & \textit{ICLR'23} & 0.0657 & 0.1111 & 2.0108 & 0.0021 & 0.0248 & 7.6848 & 0.0600 \\
\midrule
 & scGPT & \textit{Nat. Methods'24} & 0.0677 & 0.3090 & 2.7581 & 0.0038 & 0.0331 & 7.0620 & 0.0650 \\
 & scFoundation & \textit{Nat. Methods'24} & 0.0442 & 0.1681 & 4.5387 & 0.0106 & 0.0650 & 7.1529 & 0.0404 \\
Foundation & GeneCompass & \textit{Cell Res.'24} & 0.1174 & 0.2153 & 2.1901 & 0.0024 & 0.0352 & 6.9284 & 0.1118 \\
 & CellFM & \textit{Nat. Commun.'25} & 0.0634 & \best{0.3357} & 5.3577 & 0.0143 & 0.0887 & 7.6679 & 0.0432 \\
 & CellVQ & \textit{Nat. Commun.'26} & 0.0430 & 0.0518 & 2.4619 & 0.0031 & 0.0353 & 6.9286 & 0.0432 \\
\midrule
 & CPA & \textit{Mol. Syst.'23} & 0.0972 & 0.2230 & 1.9208 & 0.0019 & 0.0276 & \best{6.6606} & 0.0958 \\
 & CellFlow & \textit{bioRxiv'25} & \second{0.1437} & 0.0309 & 2.1637 & 0.0024 & \second{0.0238} & 7.0127 & \second{0.1468} \\
 & scDFM & \textit{ICLR'26} & 0.0100 & 0.2373 & 3.6133 & 0.0068 & 0.0516 & 8.3818 & 0.0025 \\
Generative & PerturbDiff & \textit{ICML'26} & -0.0300 & 0.2342 & 9.7288 & 0.0470 & 0.1123 & 11.2291 & -0.0258 \\
 & DeSCOPE & \textit{bioRxiv'26} & 0.1286 & 0.3220 & \second{1.7480} & \second{0.0015} & 0.0282 & 6.9186 & 0.0994 \\
 & STATE & \textit{Cell'26} & 0.0316 & 0.1051 & 4.1675 & 0.0087 & 0.0554 & 7.1523 & 0.0276 \\
 & \textbf{ChromaPert} & \textit{Proposed} & \best{0.2210} & \second{0.3235} & \best{1.5875} & \best{0.0013} & \best{0.0188} & \second{6.8687} & \best{0.2188} \\
\bottomrule
\end{tabular}%
}
\end{table}

\section{Experiments}
\label{sec:experiments}

\subsection{Experimental Setup}
\label{sec:experimental_setup}

\paragraph{Datasets and protocols.}
We use paired RNA--ATAC from Perturb-Multiome~\citep{martin2025transcription}
(GSE274113; 19 targets across differentiating hematopoietic states) and
CAT-ATAC~\citep{shevade2025simultaneous}
(GSE288996; 18 targets in DMSO-treated K562).
Both test unseen targets; GSE also evaluates 19 known targets across seven
held-out states (133 conditions). Supplied test cell-type counts weight
population means (Appendix~\ref{app:inference}). Data composition,
preprocessing, splits, coupling, and query controls are detailed in
Appendices~\ref{app:datasets}--\ref{app:protocol}.

\paragraph{Comparison methods.}
The 14 learned baselines comprise
(1) \textbf{graph-based models}: \textbf{GEARS},
\textbf{CellOracle}, and \textbf{GraphVCI}
~\citep{roohani2024predicting,kamimoto2023dissecting,wu2022predicting};
(2) \textbf{foundation models}: \textbf{scGPT},
\textbf{scFoundation}, \textbf{GeneCompass}, \textbf{CellFM}, and
\textbf{CellVQ}
~\citep{cui2024scgpt,hao2024large,yang2024genecompass,
zeng2025cellfm,wang2026illuminating}; and (3) \textbf{generative models}:
\textbf{CPA}, \textbf{CellFlow}, \textbf{scDFM}, \textbf{PerturbDiff},
\textbf{DeSCOPE}, and \textbf{STATE}
~\citep{lotfollahi2023predicting,klein2025cellflow,yu2026scdfm,
yuan2026perturbdiff,wu2026decoding,adduri2026predicting}. scGPT uses its native
perturbation decoder; other foundation models use frozen embeddings with
a GEARS-style adapter. Training and prediction details appear in
Appendix~\ref{app:baselines}.

\paragraph{Metrics and reporting.}
Methods share a gene panel and evaluator; unperturbed \textit{Control} is
excluded from learned-model rankings.
Pearson$_{\Delta}$ correlates condition-mean responses across genes relative
to matched controls;
Pearson$_{\Delta,\mathrm{DE}}$ uses the 20 largest absolute observed responses.
$L2$, MSE, and MAE measure mean-expression error, discriminative cosine
measures direction, and supplementary W1 evaluates exported predictions
(Appendix~\ref{app:baselines}). Scores average held-out conditions (four
decimals); see Appendices~\ref{app:metrics} and~\ref{app:multiseed} for
definitions and three-seed summaries.

\subsection{Single-cell-line unseen-target perturbation prediction}
\label{sec:k562_target_loo}

\paragraph{Fixed-background target generalization.}\label{desc:k562}
This setting isolates unseen-target generalization at a fixed cellular background.
On K562 (Table~\ref{tab:k562_comparison}), ChromaPert improves all-gene Pearson
by 53.8\% and discriminative cosine by 49.0\% over CellFlow, while leading the
learned predictors in pointwise errors. DE correlation and W1 retain specialized
leaders, so the main advantage is broad response-shape and expression accuracy.

\begin{figure}[!t]
\centering
\includegraphics[width=\linewidth]{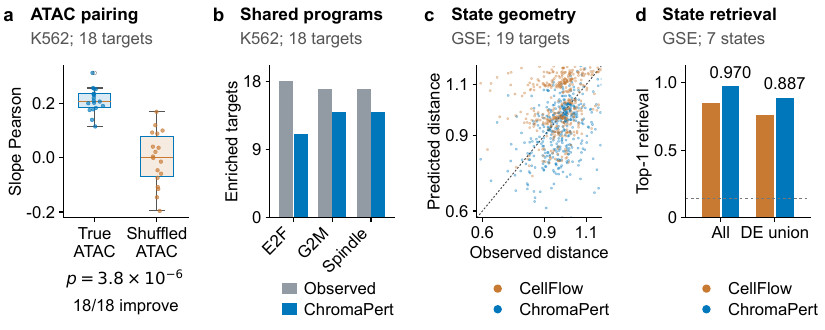}
\caption{\textbf{Chromatin-dependent and state-resolved response structure.}
\seclink{desc:chromatin}{4.2}\enspace\seclink{desc:state}{4.4}
\textbf{(a)} K562 response-slope Pearson under true/shuffled ATAC; points denote
held-out targets.
\textbf{(b)} K562 targets with enriched Hallmark programs in observed/predicted
responses (BH $q<0.05$).
\textbf{(c)} Observed/predicted GSE response distances for 19 targets with 21
state pairs each; dashed line, equality.
\textbf{(d)} Seven-state top-1 retrieval using all genes or the target-specific
DE union; dashed line, $1/7$ chance retrieval.}
\label{fig:atac_dependence}
\end{figure}

\paragraph{Chromatin-dependent biological responses.}\label{desc:chromatin}
True cell--ATAC pairing improves response-slope recovery over shuffled context
for all 18 targets (Figure~\ref{fig:atac_dependence}a): mean all-gene Pearson is
0.2120 versus $-0.0043$ (paired one-sided Wilcoxon $p=3.8\times10^{-6}$).
Both evaluations hold the frozen model and input dimensions fixed and vary
only cell--chromatin correspondence. Top-20-gene Spearman also
increases from 0.1001 to 0.2944, with improvement for 15 of 18 targets.
This diagnostic evaluates chromatin-associated response variation, not
benchmark Pearson$_\Delta$, and does not establish causality.
Hallmark enrichment~\citep{liberzon2015molecular} identifies shared
proliferative programs (Figure~\ref{fig:atac_dependence}b): G2M checkpoint and
mitotic spindle are enriched for 14 predicted versus 17 observed targets
each; E2F targets for 11 versus 18. These counts capture shared proliferative
and cell-cycle responses, rather than target-wise pathway recovery
(Appendix~\ref{app:biology}).

\subsection{Multi-cell-state unseen-target perturbation prediction}
\label{sec:gse_target_loo}

\paragraph{Multi-state target generalization.}\label{desc:gse_target}
On GSE held-out-target evaluation (Table~\ref{tab:gse_comparison}), ChromaPert
leads five of six condition-mean metrics. Relative to the strongest learned
competitors, it improves all-gene Pearson by 68.4\% and discriminative cosine by
64.0\%, while attaining the best pointwise errors and W1. GeneCompass retains
the best DE-focused correlation. Control's lower MSE but near-zero response
correlation further shows that absolute fit alone can miss response structure.

\begin{table}[!t]
\caption{Multi-cell-state unseen-target perturbation prediction on GSE.
Metrics are averaged over the 19 held-out targets. \seclink{desc:gse_target}{4.3}}
\label{tab:gse_comparison}
\centering
\scriptsize
\setlength{\tabcolsep}{2.2pt}
\renewcommand{\arraystretch}{1.08}
\resizebox{\textwidth}{!}{%
\begin{tabular}{cccccccccc}
\toprule
\multirow[c]{2}{*}{\textbf{Family}} & \multirow[c]{2}{*}{\textbf{Method}} & \multirow[c]{2}{*}{\textbf{Public.}} &
\multicolumn{2}{c}{\textbf{Mean-Response Shape}} & \multicolumn{3}{c}{\textbf{Pointwise Accuracy}} & \multicolumn{2}{c}{\textbf{Transport / Direction}} \\
\cmidrule(lr){4-5}\cmidrule(lr){6-8}\cmidrule(lr){9-10}
\textbf{} & \textbf{} & \textbf{} &
\textbf{Pearson$_{\Delta}$ $\uparrow$} &
\textbf{Pearson$_{\Delta,\mathrm{DE}}$ $\uparrow$} &
\textbf{$L2$ $\downarrow$} & \textbf{MSE $\downarrow$} &
\textbf{MAE $\downarrow$} & \textbf{W1 $\downarrow$} &
\textbf{Disc. Cos. $\uparrow$} \\
\midrule
Statistical & \textit{Control} & -- & 0.0146 & 0.0333 & 1.2374 & 0.0008 & 0.0164 & 8.8965 & 0.0132 \\
\midrule
 & GEARS & \textit{Nat. Biotechnol.'24} & 0.0263 & 0.0199 & 1.8074 & 0.0018 & 0.0248 & 7.1760 & 0.0383 \\
Graph & CellOracle & \textit{Nature'23} & 0.0897 & 0.1909 & 1.7686 & 0.0016 & 0.0235 & 10.1314 & 0.0963 \\
 & GraphVCI & \textit{ICLR'23} & 0.0073 & -0.0281 & 1.6742 & 0.0014 & 0.0223 & 10.1687 & 0.0070 \\
\midrule
 & scGPT & \textit{Nat. Methods'24} & 0.0416 & 0.0537 & 2.3209 & 0.0027 & 0.0289 & 7.3371 & 0.0436 \\
 & scFoundation & \textit{Nat. Methods'24} & 0.0275 & 0.1114 & 9.7229 & 0.1753 & 0.1444 & 10.2412 & 0.0379 \\
Foundation & GeneCompass & \textit{Cell Res.'24} & \second{0.1030} & \best{0.3843} & 6.7853 & 0.0229 & 0.1170 & 9.5319 & \second{0.1085} \\
 & CellFM & \textit{Nat. Commun.'25} & 0.0716 & 0.2386 & 7.2340 & 0.0261 & 0.1255 & 9.8175 & 0.0873 \\
 & CellVQ & \textit{Nat. Commun.'26} & 0.0059 & -0.0427 & 14.7377 & 0.1999 & 0.2162 & 12.8551 & 0.0112 \\
\midrule
 & CPA & \textit{Mol. Syst.'23} & 0.0331 & 0.1913 & 1.5268 & 0.0012 & 0.0208 & 8.8482 & 0.0113 \\
 & CellFlow & \textit{bioRxiv'25} & 0.0698 & 0.1319 & \second{1.5066} & \second{0.0012} & \second{0.0193} & \second{7.1029} & 0.0712 \\
 & scDFM & \textit{ICLR'26} & -0.0374 & -0.0513 & 3.9002 & 0.0085 & 0.0606 & 9.1254 & -0.0220 \\
Generative & PerturbDiff & \textit{ICML'26} & 0.0314 & 0.1682 & 8.9828 & 0.0402 & 0.0913 & 11.1395 & 0.0074 \\
 & DeSCOPE & \textit{bioRxiv'26} & 0.0268 & -0.0231 & 1.8125 & 0.0018 & 0.0232 & 7.2850 & -0.0214 \\
 & STATE & \textit{Cell'26} & -0.0383 & -0.0638 & 3.0031 & 0.0046 & 0.0430 & 7.3296 & -0.0594 \\
 & \textbf{ChromaPert} & \textit{Proposed} & \best{0.1735} & \second{0.3697} & \best{1.4354} & \best{0.0011} & \best{0.0186} & \best{7.1014} & \best{0.1779} \\
\bottomrule
\end{tabular}%
}
\end{table}

\subsection{Cross-cell-state perturbation transfer}
\label{sec:gse_cross_state_loo}

\paragraph{Cross-state transfer.}\label{desc:crossstate}
The GSE held-out-state setting (Table~\ref{tab:gse_cross_state_comparison})
tests transfer of known perturbations to an unseen state using its unperturbed
RNA--ATAC context. This protocol requires the model to reuse perturbation
knowledge while adapting the predicted response to a cellular background
absent from training. ChromaPert leads all six condition-mean metrics,
improving all-gene and DE-focused Pearson by 23.2\% and 46.2\%, respectively,
over the strongest learned baselines. Its lower expression errors and W1 show
that this gain does not trade off pointwise or distributional accuracy.

\begin{figure}[!t]
\centering
\includegraphics[width=\linewidth]{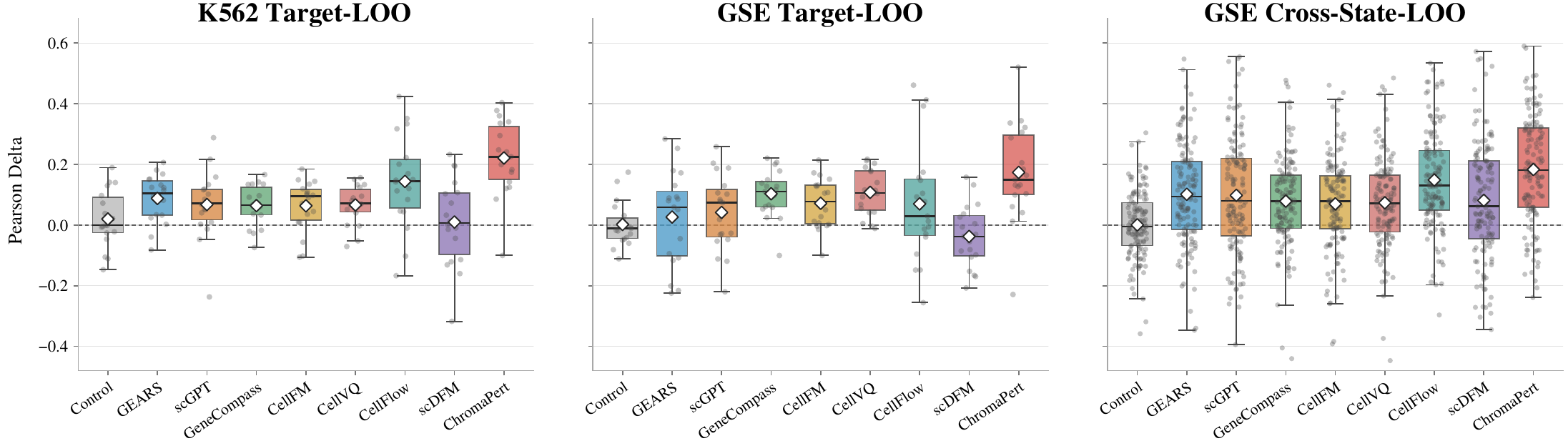}
\caption{Fold-level distribution of the response correlation
(Pearson$_{\Delta}$) across the three held-out evaluation settings. Each box
summarizes completed folds for the corresponding setting; the center line
denotes the median and whiskers show the non-outlier range. \seclink{desc:folds}{4.4}}
\label{fig:pearson_delta_boxplot}
\end{figure}

\paragraph{State-resolved response structure.}\label{desc:state}
Across the 21 state pairs per target, predicted and observed response distances
have target-mean Spearman 0.385 (Figure~\ref{fig:atac_dependence}c). Correct-state
retrieval improves over CellFlow from 0.850 to 0.970 across all genes and from
0.759 to 0.887 on the target-specific DE union (Figure~\ref{fig:atac_dependence}d),
supporting state-resolved response consistency rather than cell-type
classification (Appendix~\ref{app:biology}).
Together, these analyses test both relative geometry between states and whether
each predicted response remains closest to its matching state.
\label{desc:folds}Figure~\ref{fig:pearson_delta_boxplot} summarizes fold-level
response-correlation variability across all three evaluation settings.

\begin{table}[!t]
\caption{Cross-cell-state perturbation transfer on GSE.
Metrics are averaged over 133 held-out state--target conditions; STATE is
unavailable in this setting. \seclink{desc:crossstate}{4.4}}
\label{tab:gse_cross_state_comparison}
\centering
\scriptsize
\setlength{\tabcolsep}{2.2pt}
\renewcommand{\arraystretch}{1.08}
\resizebox{\textwidth}{!}{%
\begin{tabular}{cccccccccc}
\toprule
\multirow[c]{2}{*}{\textbf{Family}} & \multirow[c]{2}{*}{\textbf{Method}} & \multirow[c]{2}{*}{\textbf{Public.}} &
\multicolumn{2}{c}{\textbf{Mean-Response Shape}} & \multicolumn{3}{c}{\textbf{Pointwise Accuracy}} & \multicolumn{2}{c}{\textbf{Transport / Direction}} \\
\cmidrule(lr){4-5}\cmidrule(lr){6-8}\cmidrule(lr){9-10}
\textbf{} & \textbf{} & \textbf{} &
\textbf{Pearson$_{\Delta}$ $\uparrow$} &
\textbf{Pearson$_{\Delta,\mathrm{DE}}$ $\uparrow$} &
\textbf{$L2$ $\downarrow$} & \textbf{MSE $\downarrow$} &
\textbf{MAE $\downarrow$} & \textbf{W1 $\downarrow$} &
\textbf{Disc. Cos. $\uparrow$} \\
\midrule
Statistical & \textit{Control} & -- & 0.0009 & -0.0040 & 1.4238 & 0.0011 & 0.0169 & 7.0702 & 0.0008 \\
\midrule
 & GEARS & \textit{Nat. Biotechnol.'24} & 0.1013 & 0.1812 & 3.0964 & 0.0056 & 0.0439 & 7.9511 & 0.0891 \\
Graph & CellOracle & \textit{Nature'23} & 0.0999 & 0.2060 & 1.9807 & 0.0021 & 0.0244 & 7.5019 & 0.0942 \\
 & GraphVCI & \textit{ICLR'23} & 0.0906 & 0.1851 & \second{1.9190} & \second{0.0020} & \second{0.0236} & \second{7.4817} & 0.0852 \\
\midrule
 & scGPT & \textit{Nat. Methods'24} & 0.0978 & 0.1686 & 9.4233 & 0.0518 & 0.1074 & 11.6138 & 0.0971 \\
 & scFoundation & \textit{Nat. Methods'24} & 0.0479 & 0.1100 & 11.1176 & 0.0689 & 0.1631 & 10.7156 & 0.0423 \\
Foundation & GeneCompass & \textit{Cell Res.'24} & 0.0787 & 0.1555 & 7.6563 & 0.0302 & 0.1177 & 9.6319 & 0.0694 \\
 & CellFM & \textit{Nat. Commun.'25} & 0.0696 & 0.1428 & 7.9967 & 0.0327 & 0.1240 & 9.9307 & 0.0590 \\
 & CellVQ & \textit{Nat. Commun.'26} & 0.0228 & 0.0344 & 6.5700 & 0.0258 & 0.0977 & 7.8899 & 0.0241 \\
\midrule
 & CPA & \textit{Mol. Syst.'23} & 0.1236 & \second{0.2262} & 5.3573 & 0.0183 & 0.0565 & 8.7267 & 0.1215 \\
 & CellFlow & \textit{bioRxiv'25} & \second{0.1484} & 0.2197 & 4.1406 & 0.0236 & 0.0469 & 8.4091 & \second{0.1478} \\
 & scDFM & \textit{ICLR'26} & 0.0812 & 0.1312 & 9.2382 & 0.0490 & 0.1109 & 12.4967 & 0.0812 \\
Generative & PerturbDiff & \textit{ICML'26} & 0.0742 & 0.1588 & 12.4345 & 0.0825 & 0.1181 & 13.8396 & 0.0696 \\
 & DeSCOPE & \textit{bioRxiv'26} & 0.1010 & 0.1693 & 9.3057 & 0.0506 & 0.1042 & 11.6300 & 0.1006 \\
 & STATE & \textit{Cell'26} & -- & -- & -- & -- & -- & -- & -- \\
 & \textbf{ChromaPert} & \textit{Proposed} & \best{0.1829} & \best{0.3307} & \best{1.4309} & \best{0.0011} & \best{0.0173} & \best{7.0648} & \best{0.1809} \\
\bottomrule
\end{tabular}%
}
\end{table}

\raggedbottom
\vspace{5pt}
\subsection{Ablation Analysis}
\label{sec:ablations}
\vspace{2pt}

\begin{table}[!t]
\caption{ChromaPert ablations. \seclink{desc:structured}{4.5} Higher is better for
Pearson$_{\Delta}$ and Pearson$_{\Delta,\mathrm{DE}}$; lower is better for MSE
and MAE. Bold shaded rows mark the primary configuration in each panel.}
\label{tab:ablation_set}
\centering
\scriptsize
\captionsetup[subtable]{font=scriptsize,skip=1pt,labelformat=parens}
\setlength{\tabcolsep}{3.0pt}
\renewcommand{\arraystretch}{1.04}

\begin{minipage}[t]{0.49\linewidth}
\vspace{0pt}
\centering

\begin{subtable}[t]{\linewidth}
\centering
\resizebox{\linewidth}{!}{%
\begin{tabular}{lcccc}
\toprule
\textbf{Method} & Pearson$_{\Delta}\uparrow$ &
Pearson$_{\Delta,\mathrm{DE}}\uparrow$ & MSE$\downarrow$ & MAE$\downarrow$ \\
\midrule
CellFlow & 0.1484 & 0.2197 & 0.0236 & 0.0469 \\
CellFlow + ATAC & 0.0842 & 0.0841 & 0.0073 & 0.0336 \\
\rowcolor{gray!12}\textbf{ChromaPert} &
\textbf{0.1829} & \textbf{0.3307} &
\textbf{0.0011} & \textbf{0.0173} \\
\bottomrule
\end{tabular}}
\caption{ATAC-augmented control on GSE held-out-state transfer. \seclink{desc:structured}{4.5}}
\label{tab:ablation_cellflow_atac}
\end{subtable}
\end{minipage}
\hfill
\begin{minipage}[t]{0.49\linewidth}
\vspace{0pt}
\centering

\begin{subtable}[t]{\linewidth}
\centering
\resizebox{\linewidth}{!}{%
\begin{tabular}{@{}lcccc@{}}
\toprule
\textbf{Variant} & Pearson$_{\Delta}\uparrow$ &
Pearson$_{\Delta,\mathrm{DE}}\uparrow$ & MSE$\downarrow$ & MAE$\downarrow$ \\
\midrule
Direct MLP & 0.1534 & 0.3017 & 0.0011 & 0.0186 \\
w/o ATAC/locus match &
0.1393 & 0.3048 & 0.0011 & 0.0190 \\
w/o source decomposition &
0.1016 & 0.2193 & 0.0011 & 0.0190 \\
\rowcolor{gray!12}\textbf{Full} &
\textbf{0.1735} & \textbf{0.3697} & \textbf{0.0011} & \textbf{0.0186} \\
\bottomrule
\end{tabular}}
\caption{Source and flow ablations on GSE held-out-target evaluation. \seclink{desc:source}{4.5}}
\label{tab:ablation_source}
\end{subtable}

\end{minipage}

\par\vspace{0pt}

\begin{minipage}[t]{0.49\linewidth}
\vspace{0pt}
\centering

\setcounter{subtable}{2}
\begin{subtable}[t]{\linewidth}
\centering
\resizebox{\linewidth}{!}{%
\begin{tabular}{ccccc}
\toprule
\textbf{Temperature} & Pearson$_{\Delta}\uparrow$ &
Pearson$_{\Delta,\mathrm{DE}}\uparrow$ & MSE$\downarrow$ & MAE$\downarrow$ \\
\midrule
0.05 & 0.1616 & 0.3363 & 0.0011 & 0.0188 \\
0.10 & 0.1605 & 0.3359 & 0.0011 & 0.0188 \\
\rowcolor{gray!12}\textbf{0.20 (default)} &
\textbf{0.1735} & \textbf{0.3697} &
\textbf{0.0011} & \textbf{0.0186} \\
0.40 & 0.1516 & 0.3272 & 0.0011 & 0.0188 \\
0.80 & 0.1382 & 0.3091 & 0.0011 & 0.0191 \\
\bottomrule
\end{tabular}}
\caption{CGRR routing temperature on GSE held-out-target evaluation. \seclink{desc:routing}{4.5}}
\label{tab:ablation_temperature}
\end{subtable}

\end{minipage}
\hfill
\begin{minipage}[t]{0.49\linewidth}
\vspace{0pt}
\centering

\begin{subtable}[t]{\linewidth}
\centering
\resizebox{\linewidth}{!}{%
\begin{tabular}{ccccc}
\toprule
\textbf{Aggregation} & Pearson$_{\Delta}\uparrow$ &
Pearson$_{\Delta,\mathrm{DE}}\uparrow$ & MSE$\downarrow$ & MAE$\downarrow$ \\
\midrule
\rowcolor{gray!12}\textbf{Official/test} &
\textbf{0.1735} & \textbf{0.3697} &
\textbf{0.0011} & \textbf{0.0186} \\
Training-control & 0.1710 & 0.3670 & 0.0011 & 0.0187 \\
Uniform & 0.1533 & 0.3113 & 0.0011 & 0.0198 \\
\bottomrule
\end{tabular}}
\caption{Aggregation weights on GSE held-out-target evaluation. \seclink{desc:routing}{4.5}}
\label{tab:ablation_aggregation}
\end{subtable}

\end{minipage}
\end{table}

\paragraph{Structured use of ATAC.}\label{desc:structured}
Table~\ref{tab:ablation_set}(\subref{tab:ablation_cellflow_atac}) compares
ChromaPert with CellFlow and CellFlow + ATAC on 133 held-out-state conditions.
ATAC concatenation lowers Pearson$_{\Delta}$ from 0.1484 to 0.0842 and
Pearson$_{\Delta,\mathrm{DE}}$ from 0.2197 to 0.0841; ChromaPert remains best
on all four criteria. This favors context-aware organization over direct
RNA--ATAC concatenation.

\paragraph{Source decomposition and flow prediction.}\label{desc:source}
In Table~\ref{tab:ablation_set}(\subref{tab:ablation_source}), replacing the
conditional flow with Direct MLP lowers Pearson$_{\Delta}$ from 0.1735 to 0.1534
and Pearson$_{\Delta,\mathrm{DE}}$ from 0.3697 to 0.3017; removing routing or
source decomposition also lowers correlation. Flow therefore improves response
shape, while source decomposition separates control variation from perturbation
changes.

\paragraph{Routing and aggregation sensitivity.}\label{desc:routing}
Table~\ref{tab:ablation_set}(\subref{tab:ablation_temperature}) shows that
intermediate routing gives the best correlations. Training-control aggregation
retains Pearson$_{\Delta}=0.1710$ and Pearson$_{\Delta,\mathrm{DE}}=0.3670$
versus 0.1735 and 0.3697 under official weights, whereas uniform weights are
lower, supporting composition-aware routing. Further controls appear in
Appendices~\ref{app:backbone_aggregation}, \ref{app:controlled_baselines}, and
\ref{app:hyperparams}.

\paragraph{External-prior ablation.}\label{desc:prior}
\label{sec:ablation_priors}
In Table~\ref{tab:ablation_priors}, removing ESM-2 gives the largest loss
(Pearson$_{\Delta}$ 0.1735$\rightarrow$0.1569; DE 0.3697$\rightarrow$0.3268;
cosine 0.1779$\rightarrow$0.1637), identifying protein sequence as the main
target cue among the evaluated external priors. CLM-X mainly raises MSE/MAE to
0.001101/0.018953, while cCRE and Multi-TSS removal affect all-gene and DE
correlation, respectively. Pretrained-only and regulatory-only variants reach
0.1086 and 0.1585 versus 0.1735 for Full. The complementary molecular and locus
priors therefore contextualize the perturbation, with conditional flow learning
the remaining displacement (Table~\ref{tab:ablation_priors};
Appendix~\ref{app:prior_ablation}).

\begin{figure}[H]
\vspace{-5pt}
\centering
\includegraphics[width=\linewidth]{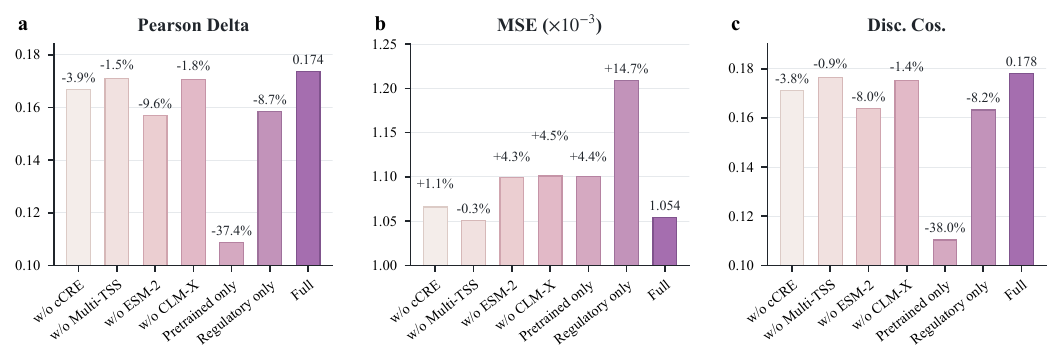}
\caption{Prior ablations on GSE target-LOO; changes are relative to Full. \seclink{desc:prior}{4.5}}
\label{fig:prior_ablation_selected}
\vspace{-5pt}
\end{figure}

\section{Conclusion}
\label{sec:conclusion}

ChromaPert separates pooled baseline offsets from target-specific perturbation
responses and conditions transport on paired control RNA--ATAC context. Its
structured target/locus conditioning and chromatin-guided routing preserve
state-dependent response structure, while conditional flow models the remaining
displacement. This decomposition lets the same perturbation acquire distinct
response representations across cellular contexts without treating
pre-existing control differences as perturbation effects. Together, these
components support expression prediction for unseen perturbations and transfer
of known perturbations to new cellular states within a common formulation for
learning from unpaired cell populations, making measured chromatin context
directly actionable for response prediction across diverse cellular backgrounds
and targets.

\clearpage
\section*{Reproducibility Statement}
Section~\ref{sec:method} describes the model formulation, training target,
inference procedure, and information boundaries. Dataset provenance,
preprocessing, fixed evaluation splits, prediction-time control pools,
response construction, aggregation, and metric definitions are documented in
Appendices~\ref{app:datasets}--\ref{app:metrics}. Comparator interfaces are
described in Appendix~\ref{app:baselines}; prior construction and numerical
flow details appear in Appendices~\ref{app:priors}--\ref{app:flow}; and the
architecture, optimization settings, and primary hyperparameters are given
in Appendix~\ref{app:implementation}. Repeat-run summaries and controlled
ablations are reported in Appendices~\ref{app:multiseed},
\ref{app:controlled_baselines}, and~\ref{app:hyperparams}. We will release
the implementation code, model checkpoints, and documentation separately.
All datasets used in this work are publicly
available through their original repositories, with accession information in
Appendix~\ref{app:datasets}.

\section*{Ethics Statement}
This work is a secondary computational analysis of publicly available
single-cell datasets. We recruited no new human participants, performed no
clinical intervention, and accessed no direct personal identifiers. Raw data
are not redistributed; access and reuse remain subject to the terms of the
original studies and repositories. ChromaPert is intended for computational
research and hypothesis generation, not for clinical diagnosis or treatment
selection. Its predictions may inherit technical or biological biases from
the source assays, cell-state annotations, and external reference resources,
and therefore require independent experimental validation before biomedical
use.

\section*{AI Use Statement}
Generative AI tools assisted manuscript editing and organization, discussion
of author-supplied results, and figure preparation. The authors are responsible
for checking AI-assisted content against the underlying data, code, and
references, and for the final text, claims, equations, figures, and results.

\bibliography{references/google_scholar_raw}
\bibliographystyle{iclr2027_conference}

\clearpage
\appendix
\raggedbottom
\section{Supplementary Experiments}
\label{app:experiments}
\setcounter{figure}{0}
\renewcommand{\thefigure}{A\arabic{figure}}

\subsection{Direct regression and aggregation controls}
\label{app:backbone_aggregation}

\paragraph{Matched residual-prediction task.}
We compare ChromaPert with Direct MLP on all 19 GSE held-out-target folds using
the configuration in Table~\ref{tab:gse_comparison}. Both retain the same
target/locus and control-state conditions, CGRR source decomposition, and PCA
residual target. This comparison isolates the contribution of flow relative
to direct regression under the same source-estimation procedure.

\paragraph{Aggregation and evaluation.}
Three rules aggregate state-specific responses: \textbf{Official/test} uses each
target's supplied full test cell-type counts; \textbf{Training-control} uses
training-control frequencies; and \textbf{Uniform} weights the aggregation
groups equally. Group support, the matched-control baseline, test cells, and
the evaluator remain fixed. After response aggregation and baseline addition,
each condition mean is exported as 32 identical rows. The row count is an
export convention, not the number of observed test cells.

\begin{table}[!htbp]
\caption{Flow versus direct regression under three aggregation rules on GSE
held-out-target evaluation (19 targets). Bold marks the better value per pair; displayed ties
are both bold.}
\label{tab:app_backbone_aggregation}
\centering
\scriptsize
\setlength{\tabcolsep}{4pt}
\renewcommand{\arraystretch}{1.10}
\resizebox{\linewidth}{!}{%
\begin{tabular}{llrrrrr}
\toprule
\textbf{Aggregation} & \textbf{Backbone} &
Pearson$_{\Delta}\uparrow$ & Pearson$_{\Delta,\mathrm{DE}}\uparrow$ &
$L2\downarrow$ & MSE$\downarrow$ & MAE$\downarrow$ \\
\midrule
\multirow{2}{*}{Official/test}
 & Direct MLP & 0.153359 & 0.301672 & 1.437177 & 0.001057 & \textbf{0.018598} \\
 & ChromaPert & \textbf{0.173501} & \textbf{0.369660} & \textbf{1.435411} & \textbf{0.001054} & 0.018632 \\
\midrule
\multirow{2}{*}{Training-control}
 & Direct MLP & 0.150976 & 0.298344 & 1.439621 & 0.001060 & \textbf{0.018680} \\
 & ChromaPert & \textbf{0.170987} & \textbf{0.366970} & \textbf{1.438524} & \textbf{0.001058} & 0.018739 \\
\midrule
\multirow{2}{*}{Uniform}
 & Direct MLP & 0.138576 & 0.263558 & \textbf{1.465007} & \textbf{0.001096} & \textbf{0.019482} \\
 & ChromaPert & \textbf{0.153270} & \textbf{0.311312} & 1.465979 & \textbf{0.001096} & 0.019792 \\
\bottomrule
\end{tabular}}
\end{table}

\paragraph{Flow improves response correlations beyond source estimation.}
ChromaPert has higher Pearson$_{\Delta}$ and Pearson$_{\Delta,\mathrm{DE}}$
than Direct MLP under every aggregation rule. With official weights,
the relative gains are 13.1\% and 22.5\%, respectively. Absolute errors
are close: Direct MLP has slightly lower MAE in all three comparisons and
slightly lower $L2$ under uniform weights. Flow's advantage in this comparison
is therefore concentrated in the cross-gene mean-response pattern, with no
uniform improvement in absolute error.

\paragraph{Sensitivity to composition weights.}
Training-control aggregation retains 98.6\% of ChromaPert's official
Pearson$_{\Delta}$ and 99.3\% of its Pearson$_{\Delta,\mathrm{DE}}$.
Uniform aggregation lowers these correlations to 0.153270 and 0.311312,
showing sensitivity to population weighting. Training-control frequencies
retain most of the correlation advantage, and flow has higher response
correlations than MLP under all three rules. These comparisons vary only the
response weights: group support and the composition-weighted control
reference remain fixed.

\subsection{Biological response and chromatin-state audits}
\label{app:biology}

\begin{figure}[!htbp]
\centering
\includegraphics[width=\linewidth]{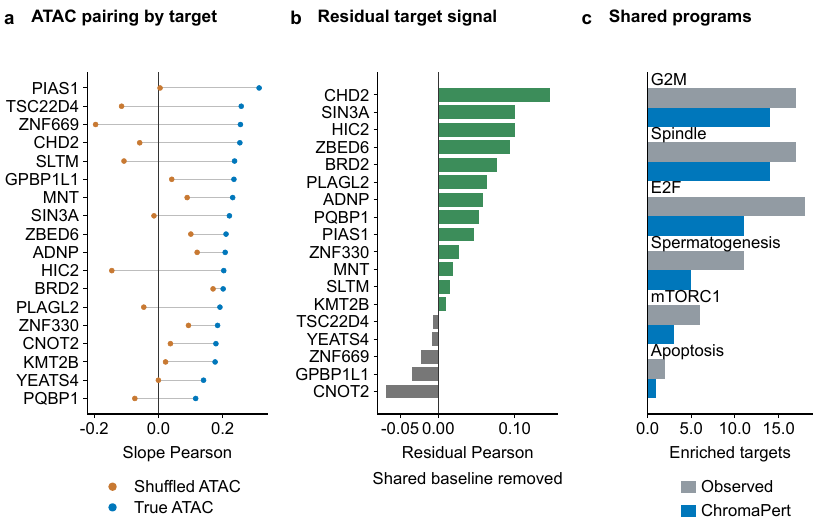}
\caption{\textbf{Extended K562 response audits.}
\textbf{(a)} Target-wise response-slope Pearson under true and shuffled ATAC
pairing, with lines connecting results for the same target.
\textbf{(b)} Residual target signal after removing the shared response
baseline. Positive and negative values are shown in green and gray.
\textbf{(c)} The expanded set of shared response programs displayed in the
state-discovery analysis. Bars count enriched targets in observed and
predicted responses. E2F, G2M, and spindle abbreviate E2F targets, G2M
checkpoint, and mitotic spindle.}
\label{fig:app_k562_biology}
\end{figure}

\paragraph{Chromatin pairing and shared functional programs.}
For each of the 18 held-out K562 targets, the response-slope diagnostic
compares true control cell--chromatin pairing against ATAC context shuffled
across cells. Both conditions use the same frozen model and input dimensions.
All-gene slope Pearson averages 0.2120 with true pairing and $-0.0043$ after
shuffling, with an improvement for every target and a paired one-sided
Wilcoxon $p=3.81\times10^{-6}$.
Top-20-gene Spearman averages 0.2944 versus 0.1001, with improvements for
15 of 18 targets. These statistics quantify recovery of chromatin-associated
response slopes, whereas benchmark Pearson$_\Delta$ measures condition-mean
expression changes.
Figure~\ref{fig:app_k562_biology}a shows the full target-wise comparison.
The accompanying context-swap audit reports nonzero prediction changes for
all 18 targets, averaging $9.080\times10^{-3}$ in absolute magnitude.

Hallmark gene sets~\citep{liberzon2015molecular} summarize recurring functional
programs in observed and predicted responses.
Figure~\ref{fig:app_k562_biology}c expands the three proliferation-related
programs shown in the main figure. Predictions contain G2M-checkpoint and
mitotic-spindle enrichment for 14 targets each, compared with 17 in the
observations, and E2F enrichment for 11 versus 18. The original pathway audit
also reports mTORC1 enrichment for three predicted versus six observed
targets, and apoptosis and p53 enrichment for one versus two each.
These counts summarize how often each program is enriched; they do not
measure agreement for individual targets.

\paragraph{Candidate target-specific signals.}
After shared-baseline removal, CHD2, SIN3A, HIC2, ZBED6, and BRD2 have the
largest positive residual signals in the target audit
(Figure~\ref{fig:app_k562_biology}b).
The ranking identifies candidate chromatin-sensitive perturbations for
follow-up. It remains exploratory because target-label permutation provides
limited support for pathway specificity. The functional interpretation
therefore rests primarily on the shared proliferation programs.

\begin{figure}[!htbp]
\centering
\includegraphics[width=\linewidth]{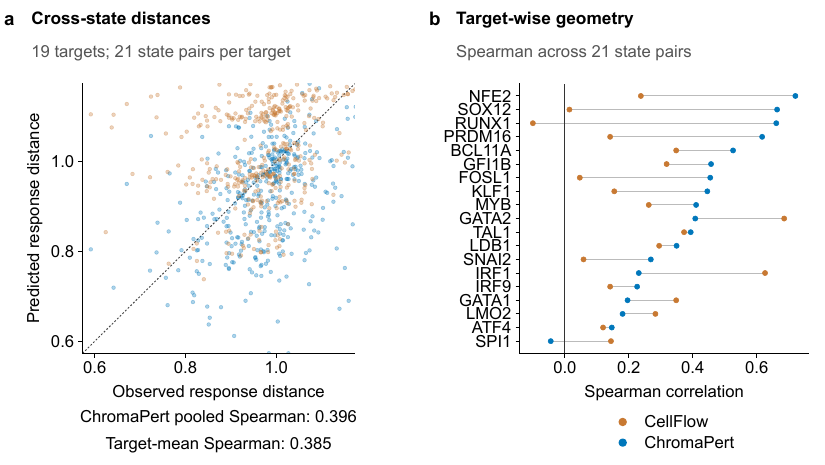}
\caption{\textbf{GSE held-out-state response geometry at target resolution.}
\textbf{(a)} Observed and predicted response distances, with each point
representing one of 21 state pairs for a target. The dashed diagonal denotes
equal distance.
\textbf{(b)} Spearman correlation over the 21 pairs, shown separately for all
19 targets and both methods. Lines connect the same target, and the vertical
reference marks zero correlation.}
\label{fig:app_gse_geometry}
\end{figure}

\paragraph{Held-out-state response geometry and retrieval.}
The GSE analysis uses the seven annotated states from held-out-state evaluation.
For each of 19 targets, all $\binom{7}{2}=21$ state pairs contribute an
observed and a predicted response distance.
Figure~\ref{fig:app_gse_geometry} reports the pooled comparison and the
target-wise Spearman correlations. ChromaPert's mean target-wise Spearman is
0.385 and its pooled Spearman is 0.396. The per-target results vary
across perturbations, and CellFlow has the higher correlation for some targets.
We report the target-stratified summary because pairs sharing a
state or target are dependent.

The retrieval audit matches each predicted state--target response against
the seven observed state responses for that same target. All-gene top-1
retrieval is 0.970 for ChromaPert and 0.850 for CellFlow. Restricting the
comparison to the target-specific DE union gives 0.887 and 0.759,
respectively. The reference level for uniform retrieval among seven states
is $1/7$. Retrieval measures retention of state information in predicted
responses, complementing the comparison of relative distances.

\begin{figure}[!htbp]
\centering
\includegraphics[width=\linewidth]{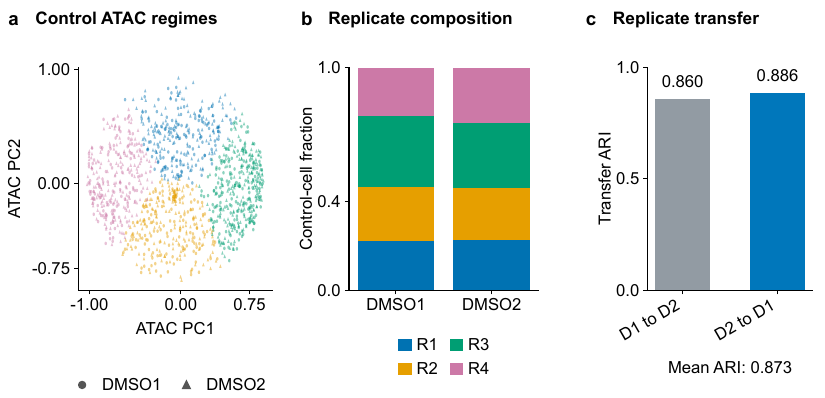}
\caption{\textbf{Candidate chromatin regimes in control K562 cells.}
\textbf{(a)} PCA display of the normalized 32-axis ATAC representation, colored
by the four KMeans regimes. Circles and triangles distinguish DMSO1 and DMSO2.
\textbf{(b)} Regime proportions within each control replicate.
\textbf{(c)} Bidirectional replicate-transfer ARI. D1 and D2 denote DMSO1 and
DMSO2; ARI measures partition agreement. Clustering is a post-hoc input audit
with $k=4$, while ChromaPert conditions on continuous chromatin features.}
\label{fig:app_chromatin_regimes}
\end{figure}

\paragraph{Label-free control-chromatin variation.}
The state-discovery analysis is a post-hoc audit of control chromatin, carried
out without retraining ChromaPert. It uses a control-fitted, technically
residualized 32-axis ATAC representation combining fixed-bin, cCRE,
gene-activity, and motif-activity views. Of the 5,412 metadata-aligned cells
in this analysis subset, 1,261 are controls: 547 from DMSO1 and 714 from DMSO2.
These counts describe the audit subset, which is distinct from the benchmark
partitions in Table~\ref{tab:app_datasets}.
After unit-norm normalization, KMeans with the exploratory choice $k=4$
defines four candidate chromatin regimes using control ATAC alone.
Cell-state annotations and post-perturbation ATAC are excluded from this
clustering step. Figure~\ref{fig:app_chromatin_regimes} shows the candidate
partitions, their representation in both replicates, and bidirectional
replicate-transfer agreement. Transfer ARI is 0.860 from DMSO1 to DMSO2 and
0.886 in the reverse direction, averaging 0.873. This agreement across
replicates supports using control-chromatin variation as context for response
prediction.

\subsection{Repeat-run performance}
\label{app:multiseed}

Table~\ref{tab:app_multiseed} summarizes protocol-level scores across three
random seeds as means and standard deviations. These summaries complement
the primary evaluation tables in the main text. The percentage gains in the
abstract and introduction use the primary evaluation tables, not these
three-seed means. The standard deviations describe between-run variation;
they are not confidence intervals over held-out targets or states.

\begingroup
\setlength{\intextsep}{4pt}
\begin{table}[!htbp]
\caption{Three-seed comparison of CellFlow and ChromaPert (mean and SD).
Bold marks the better mean.}
\label{tab:app_multiseed}
\centering
\footnotesize
\setlength{\tabcolsep}{6pt}
\renewcommand{\arraystretch}{1.00}
\captionsetup[subtable]{font=footnotesize,skip=2pt,labelformat=parens}

\begin{subtable}{0.96\linewidth}
\centering
\begin{tabular*}{\linewidth}{@{\extracolsep{\fill}}lrrrr@{}}
\toprule
\multirow{2}{*}{\textbf{Metric}} &
\multicolumn{2}{c}{\textbf{CellFlow}} &
\multicolumn{2}{c}{\textbf{ChromaPert}} \\
\cmidrule(lr){2-3}\cmidrule(lr){4-5}
 & Mean & SD & Mean & SD \\
\midrule
Pearson$_{\Delta}$ $\uparrow$ &
0.1038 & 0.0117 & \textbf{0.230283} & 0.001164 \\
Pearson$_{\Delta,\mathrm{DE}}$ $\uparrow$ &
0.0486 & 0.0173 & \textbf{0.325833} & 0.006795 \\
\addlinespace[3pt]
$L2$ $\downarrow$ &
2.1344 & 0.0715 & \textbf{1.588487} & 0.000941 \\
MSE $\downarrow$ &
0.002318 & 0.000153 & \textbf{0.001289} & 0.000001 \\
MAE $\downarrow$ &
0.02363 & 0.00064 & \textbf{0.018822} & 0.000005 \\
\addlinespace[3pt]
W1 $\downarrow$ &
7.0740 & 0.0179 & \textbf{6.869656} & 0.000102 \\
Disc. Cos. $\uparrow$ &
0.1047 & 0.0140 & \textbf{0.228007} & 0.000967 \\
\bottomrule
\end{tabular*}
\caption{K562 target-LOO (18 targets).}
\label{tab:app_multiseed_k562}
\end{subtable}

\par\vspace{1.5mm}

\begin{subtable}{0.96\linewidth}
\centering
\begin{tabular*}{\linewidth}{@{\extracolsep{\fill}}lrrrr@{}}
\toprule
\multirow{2}{*}{\textbf{Metric}} &
\multicolumn{2}{c}{\textbf{CellFlow}} &
\multicolumn{2}{c}{\textbf{ChromaPert}} \\
\cmidrule(lr){2-3}\cmidrule(lr){4-5}
 & Mean & SD & Mean & SD \\
\midrule
Pearson$_{\Delta}$ $\uparrow$ &
0.071317 & 0.008266 & \textbf{0.158009} & 0.017618 \\
Pearson$_{\Delta,\mathrm{DE}}$ $\uparrow$ &
0.147989 & 0.036863 & \textbf{0.334136} & 0.039319 \\
\addlinespace[3pt]
$L2$ $\downarrow$ &
1.501639 & 0.007458 & \textbf{1.455663} & 0.022097 \\
MSE $\downarrow$ &
0.001145 & 0.000010 & \textbf{0.001083} & 0.000039 \\
MAE $\downarrow$ &
0.019312 & 0.000058 & \textbf{0.018831} & 0.000222 \\
\addlinespace[3pt]
W1 $\downarrow$ &
\textbf{7.101689} & 0.001898 & 7.105784 & 0.005875 \\
Disc. Cos. $\uparrow$ &
0.073415 & 0.008620 & \textbf{0.162550} & 0.017430 \\
\bottomrule
\end{tabular*}
\caption{GSE target-LOO (19 targets).}
\label{tab:app_multiseed_gse_target}
\end{subtable}

\par\vspace{1.5mm}

\begin{subtable}{0.96\linewidth}
\centering
\begin{tabular*}{\linewidth}{@{\extracolsep{\fill}}lrrrr@{}}
\toprule
\multirow{2}{*}{\textbf{Metric}} &
\multicolumn{2}{c}{\textbf{CellFlow}} &
\multicolumn{2}{c}{\textbf{ChromaPert}} \\
\cmidrule(lr){2-3}\cmidrule(lr){4-5}
 & Mean & SD & Mean & SD \\
\midrule
Pearson$_{\Delta}$ $\uparrow$ &
0.0881 & 0.0048 & \textbf{0.1757} & 0.0010 \\
Pearson$_{\Delta,\mathrm{DE}}$ $\uparrow$ &
0.0926 & 0.0156 & \textbf{0.3251} & 0.0016 \\
\addlinespace[3pt]
$L2$ $\downarrow$ &
3.0498 & 0.0580 & \textbf{1.4323} & 0.0003 \\
MSE $\downarrow$ &
0.007033 & 0.000395 & \textbf{0.001101} & 0.000001 \\
MAE $\downarrow$ &
0.03312 & 0.00059 & \textbf{0.01727} & 0.00001 \\
\addlinespace[3pt]
W1 $\downarrow$ &
7.4933 & 0.0304 & \textbf{7.0825} & 0.0002 \\
Disc. Cos. $\uparrow$ &
0.0883 & 0.0049 & \textbf{0.1737} & 0.0011 \\
\bottomrule
\end{tabular*}
\caption{GSE cross-state-LOO (133 state--target conditions).}
\label{tab:app_multiseed_gse_cross}
\end{subtable}
\end{table}

Table~\ref{tab:app_multiseed}(\subref{tab:app_multiseed_k562}) shows higher
mean-response correlations and discriminative cosine, together with lower
$L2$, MSE, and MAE, for ChromaPert than CellFlow on K562. The same pattern holds
on GSE target-LOO (\subref{tab:app_multiseed_gse_target}) and cross-state-LOO
(\subref{tab:app_multiseed_gse_cross}). ChromaPert also has lower supplementary
W1 on K562 and cross-state transfer; CellFlow retains a slightly lower W1 on
GSE target-LOO.
\par\endgroup

\subsection{External-prior ablation}
\label{app:prior_ablation}
Table~\ref{tab:ablation_priors} reports the complete external-prior ablation.
\begin{table}[!htbp]
\caption{External-prior ablation on GSE target-LOO.}
\label{tab:ablation_priors}
\centering\scriptsize
\resizebox{\textwidth}{!}{%
\begin{tabular}{lccccccc}
\toprule
\textbf{Variant} & $\mathrm{Pearson}_{\Delta}$ & $\mathrm{Pearson}_{\Delta,\mathrm{DE}}$ & $L2$ & MSE & MAE & W1 & Disc. Cos.\\
\midrule
Full & 0.173501 & 0.369660 & 1.4354 & 0.001054 & 0.018632 & 7.1014 & 0.177905\\
w/o cCRE & 0.166821 & 0.373276 & 1.4458 & 0.001066 & 0.018758 & 7.1042 & 0.171148\\
w/o Multi-TSS & 0.170940 & 0.351339 & 1.4328 & 0.001051 & 0.018580 & 7.0987 & 0.176367\\
w/o ESM-2 & 0.156857 & 0.326840 & 1.4650 & 0.001099 & 0.018886 & 7.1079 & 0.163678\\
w/o CLM-X & 0.170435 & 0.361918 & 1.4661 & 0.001101 & 0.018953 & 7.1096 & 0.175327\\
Pretrained priors only & 0.108649 & 0.248848 & 1.4674 & 0.001100 & 0.019008 & 7.1028 & 0.110360\\
Regulatory priors only & 0.158469 & 0.316061 & 1.5222 & 0.001209 & 0.019461 & 7.1274 & 0.163237\\
\bottomrule
\end{tabular}}
\end{table}

\paragraph{Protein and RNA target representations.}
ESM-2 removal reduces Pearson$_{\Delta}$ from 0.173501 to 0.156857 and
Pearson$_{\Delta,\mathrm{DE}}$ from 0.369660 to 0.326840, the largest
correlation losses among single-prior removals. Protein language models
encode sequence-derived structural information~\citep{lin2023evolutionary}.
These representations connect held-out protein targets to training targets with
similar molecular properties.
CLM-X removal has a smaller effect on these correlations but raises MSE from
0.001054 to 0.001101 and MAE from 0.018632 to 0.018953. Its frozen RNA-token
representation provides complementary target information for predicting the
magnitude of expression changes.

\paragraph{Regulatory-locus and promoter-sequence context.}
Removing cCRE decreases all-gene correlation from 0.173501 to 0.166821 and
discriminative cosine from 0.177905 to 0.171148, while DE correlation rises
slightly to 0.373276. cCRE annotations describe candidate regulatory
elements~\citep{encode2020expanded}. These results show that target-locus context
contributes to gene-wide response prediction, with a different trade-off on the
top-response subset. Multi-TSS removal lowers DE correlation
to 0.351339 while slightly improving MSE and MAE. TSS-centered sequence
windows represent alternative promoter contexts associated with a target
~\citep{abugessaisa2019reftss}. Their value may lie in distinguishing the
relative responses of strongly affected genes rather than minimizing global
expression error. These metric-specific effects motivate combining
protein-level target information with locus-level regulatory context.

\paragraph{Complementarity across prior families.}
Pretrained-only and regulatory-only variants reach all-gene correlations of
0.108649 and 0.158469, respectively, below 0.173501 for Full. Their DE
correlations and directional agreement are also lower. This advantage shows
the benefit of combining molecular target descriptors with regulatory context.
The individual ablations show which prediction criteria are most
sensitive to each information source.

\subsection{Controlled input and architecture baselines}
\label{app:controlled_baselines}

\begin{table}[!htbp]
\caption{Controlled input and architecture baselines on GSE target-LOO.
Lower $L2$, MSE, MAE, and W1 are better; higher correlations and
discriminative cosine are better.}
\label{tab:app_controlled_baselines}
\centering
\scriptsize
\setlength{\tabcolsep}{3.2pt}
\renewcommand{\arraystretch}{1.08}
\resizebox{\textwidth}{!}{%
\begin{tabular}{lccccccc}
\toprule
\textbf{Method} & \textbf{Pearson$_{\Delta}$ $\uparrow$} &
\textbf{Pearson$_{\Delta,\mathrm{DE}}$ $\uparrow$} &
\textbf{$L2$ $\downarrow$} & \textbf{MSE $\downarrow$} &
\textbf{MAE $\downarrow$} & \textbf{W1 $\downarrow$} &
\textbf{Disc. Cos. $\uparrow$} \\
\midrule
Concat-Linear-ATAC & -0.018874 & -0.053584 & 1.895157 & 0.001942 & 0.024209 & 7.305901 & -0.013298 \\
Concat-Ridge-ATAC & -0.014108 & -0.041709 & 1.889544 & 0.001929 & 0.024143 & 7.304001 & -0.008318 \\
Concat-MLP-ATAC & 0.015239 & -0.017022 & 2.132850 & 0.002410 & 0.027351 & 8.050125 & 0.032644 \\
RNA-only MLP & 0.038374 & 0.026677 & 2.034526 & 0.002155 & 0.026057 & 7.776974 & 0.049617 \\
Additive mean shift & 0.006048 & 0.030176 & 1.715645 & 0.001614 & 0.022049 & 7.269627 & 0.015198 \\
\rowcolor{gray!12}\textbf{ChromaPert} & \textbf{0.173501} & \textbf{0.369660} &
\textbf{1.435411} & \textbf{0.001054} & \textbf{0.018632} &
\textbf{7.101355} & \textbf{0.177905} \\
\bottomrule
\end{tabular}}
\end{table}

These controls compare structured chromatin conditioning with
direct ATAC concatenation, RNA-only regression, and an additive mean-shift
control. The unperturbed Control reference is reported in the main benchmark
(Table~\ref{tab:gse_comparison}).
ChromaPert leads on all six condition-mean metrics, combining higher response
correlations with lower absolute errors. It also has the lowest supplementary
W1 among these controls.

\subsection{Numerical sensitivity and inference cost}
\label{app:hyperparams}

\paragraph{ODE integration steps.}
\begin{table}[!htbp]
\caption{NFE accuracy--latency trade-off on GSE target-LOO.}
\label{tab:app_nfe}
\centering
\scriptsize
\setlength{\tabcolsep}{4.2pt}
\renewcommand{\arraystretch}{1.08}
\resizebox{0.88\textwidth}{!}{%
\begin{tabular}{cccccc}
\toprule
\textbf{NFE} & \textbf{Pearson$_{\Delta}$ $\uparrow$} &
\textbf{Pearson$_{\Delta,\mathrm{DE}}$ $\uparrow$} &
\textbf{MSE $\downarrow$} & \textbf{MAE $\downarrow$} &
\textbf{Query (s) $\downarrow$} \\
\midrule
5  & 0.173494 & 0.369639 & 0.001054 & 0.018632 & 0.004764 \\
\rowcolor{gray!12}\textbf{8 (default)} &
\textbf{0.173493} & \textbf{0.369657} & \textbf{0.001054} &
\textbf{0.018632} & \textbf{0.006947} \\
10 & 0.173497 & 0.369667 & 0.001054 & 0.018631 & 0.008818 \\
20 & 0.173500 & 0.369691 & 0.001053 & 0.018631 & 0.016500 \\
50 & 0.173500 & 0.369703 & 0.001053 & 0.018631 & 0.036394 \\
\bottomrule
\end{tabular}}
\end{table}

Table~\ref{tab:app_nfe} varies the number of function evaluations from 5 to
50. Predictive metrics change little, whereas mean
branch-query time rises from 0.004764 to 0.036394 seconds. We therefore retain
NFE$=8$: within the tested range, additional ODE steps increase latency without
a material accuracy gain.

\paragraph{Transport-mean blend.}
Table~\ref{tab:ablation_blend} reports blend sensitivity on GSE target-LOO.
The primary configuration uses a blend
of 0.18. Across the tested settings, absolute errors remain similar,
whereas mean-response correlations vary more noticeably.

\begin{table}[!htbp]
\caption{Transport-mean blend on GSE target-LOO. The bold shaded row marks
the frozen default configuration.}
\label{tab:ablation_blend}
\centering
\footnotesize
\setlength{\tabcolsep}{6pt}
\renewcommand{\arraystretch}{1.08}
\begin{tabular}{lcccc}
\toprule
\textbf{Blend} & Pearson$_{\Delta}\uparrow$ &
Pearson$_{\Delta,\mathrm{DE}}\uparrow$ & MSE$\downarrow$ & MAE$\downarrow$ \\
\midrule
0.00 & 0.1573 & 0.3366 & 0.0011 & 0.0189 \\
0.09 & 0.1581 & 0.3355 & 0.0011 & 0.0190 \\
\rowcolor{gray!12}\textbf{0.18 (default)} &
\textbf{0.1735} & \textbf{0.3697} &
\textbf{0.0011} & \textbf{0.0186} \\
0.36 & 0.1588 & 0.3391 & 0.0011 & 0.0187 \\
\bottomrule
\end{tabular}
\end{table}

\section{Data and Evaluation Protocols}
\label{app:data_evaluation}
\subsection{Dataset details and preprocessing}
\label{app:datasets}
We use the dataset names from the original studies.
\emph{GSE274113 Perturb-Multiome}
~\citep{martin2025transcription} (GEO accession GSE274113) is a paired
RNA--ATAC--guide assay of primary human
hematopoietic cells undergoing in-vitro erythroid differentiation.  The
collection includes days 7, 9, 11, and 14 and 14 biological replicates.  The
benchmark retains the non-targeting (NT) control and 19 target labels:
\texttt{ATF4}, \texttt{BCL11A}, \texttt{FOSL1}, \texttt{GATA1},
\texttt{GATA2}, \texttt{GFI1B}, \texttt{IRF1}, \texttt{IRF9}, \texttt{KLF1},
\texttt{LDB1}, \texttt{LMO2}, \texttt{MYB}, \texttt{NFE2}, \texttt{PRDM16},
\texttt{RUNX1}, \texttt{SNAI2}, \texttt{SOX12}, \texttt{SPI1}, and
\texttt{TAL1}.  Metadata contain 137,604 non-empty cells, with 32 fine-grained
cell types and ten consolidated labels.

\emph{GSE288996 CAT-ATAC K562}~\citep{shevade2025simultaneous} (GEO accession
GSE288996) is a paired K562 CAT-ATAC perturbation assay.  Although the raw accession contains iPSC,
Dasatinib, and DMSO groups, the benchmark uses only the two biological
replicates of K562 DMSO.  The strict model subset contains approximately
11,928 cells (8,347 targeting and 3,581 non-targeting) and 18 targets:
\texttt{ADNP}, \texttt{BRD2}, \texttt{CHD2}, \texttt{CNOT2},
\texttt{GPBP1L1}, \texttt{HIC2}, \texttt{KMT2B}, \texttt{MNT}, \texttt{PIAS1},
\texttt{PLAGL2}, \texttt{PQBP1}, \texttt{SIN3A}, \texttt{SLTM},
\texttt{TSC22D4}, \texttt{YEATS4}, \texttt{ZBED6}, \texttt{ZNF330}, and
\texttt{ZNF669}.  No cell-type label is used for this setting.

\begin{table}[!htbp]
\caption{Model-ready datasets and stored partitions. Protocol-specific
leave-one-out masks are described separately below.}
\label{tab:app_datasets}
\centering
\footnotesize
\setlength{\tabcolsep}{3pt}
\renewcommand{\arraystretch}{1.15}
\begin{tabularx}{\linewidth}{@{}>{\raggedright\arraybackslash}Xrrrrrr>{\raggedright\arraybackslash}p{0.15\linewidth}@{}}
\toprule
\multirow{2}{*}{\textbf{Dataset}} &
\multicolumn{3}{c}{\textbf{Features}} &
\multicolumn{3}{c}{\textbf{Fixed partitions (cells)}} &
\multirow{2}{*}{\textbf{Protocols}} \\
\cmidrule(lr){2-4}\cmidrule(lr){5-7}
 & RNA & ATAC & Total & Train & Val. & Test & \\
\midrule
Perturb-Multiome\newline GSE274113 &
2,013 & 2,013 & 4,026 & 107,471 & 11,798 & 18,335 &
Target-LOO;\newline cross-state-LOO \\
\addlinespace[4pt]
CAT-ATAC K562\newline GSE288996 (DMSO) &
2,018 & 2,018 & 4,036 & 4,063 & 878 & 471 &
Target-LOO \\
\bottomrule
\end{tabularx}
\end{table}

RNA and promoter ATAC counts are independently normalized to a library size
of 10,000 and transformed with \texttt{log1p}.
The model-ready files use a single normalization stage and retain raw library
sizes, dataset-specific scalers, and the canonical gene order in the manifest.
Gene-aligned promoter accessibility is stored with the
\texttt{ATAC\_PROM:} prefix, while RNA columns use the \texttt{RNA:} prefix.
The benchmark uses 2,013 or 2,018 promoter coordinates. Extra genes are ignored
by the adapter. Missing canonical genes must be resolved by the fixed mapping
before a method is scored.

The three evaluation settings use fixed masks. GSE target-LOO has
19 folds; each fold trains on NT controls and the other 18 targets and scores
the held-out target. Across folds, evaluation covers 19 conditions and 2,432
observed test cells (19 $\times$ 128). K562 target-LOO has 18 folds; each fold
trains on non-targeting controls and the other 17 targets. Evaluation covers
2,122 observed test cells in total. CNOT2, PLAGL2, and PQBP1 retain 53, 80, and 69 truth cells,
respectively, because fewer cells pass the strict filter.  GSE cross-state-LOO
holds out seven erythroid states, producing $7\times19=133$ state--target
conditions and 16,833 common-evaluation cells.  The held-out states are
\texttt{BFU-E}, \texttt{Basophilic Erythroblast}, \texttt{CFU-E},
\texttt{EoBasoMast Precursor}, \texttt{Orthochromatic Erythroblast},
\texttt{Polychromatic Erythroblast}, and \texttt{Pro-Erythroblast}.
MEP, megakaryocyte, and myeloid precursor labels are excluded to define the
fixed seven-state trajectory.

All methods use a common evaluation manifest specifying the same
condition list, canonical gene order, truth cells, control reference,
residual definition, metric implementation, and random seed.  In GSE
target-LOO, the prediction-time input contains a fixed pool of 1,024 control
cells and the evaluator uses frozen cell-type-matched clean pools (up to 2,048
cells per pool) to compute residuals. Cross-state transfer may use clean controls
from the held-out state only at prediction time.  K562 uses one fixed pool of
1,024 global DMSO controls for both prediction and evaluation.  These objects
are described separately from the training split in Appendix~\ref{app:protocol}.

\subsection{Fold-restricted protocol}
\label{app:protocol}
The model is fitted separately within each fixed split. The promoter mapping
$M$, cell-wise library-size transforms, canonical gene order, and fixed sequence,
TSS, cCRE, and motif resources are defined before splitting. ATAC projections,
prototypes, pooled control references, control-state offsets, source mixtures, and
response scales are recomputed from the training portion of each fold.
All trainable parameters are estimated from this portion alone.

For a state-indexed setting, the training coupling is
\begin{equation}
 \pi_{s,p}=D_0(s)\otimes D_1(s,p),
 \qquad
 \pi_p=\sum_{s\in\mathcal{S}_p}\omega_{s\mid p}\pi_{s,p},
 \qquad
 \omega_{s\mid p}=\frac{n_{s,p}}{\sum_{s'}n_{s',p}},
 \label{eq:app_coupling}
\end{equation}
where $D_0(s)$ and $D_1(s,p)$ are the training control and perturbed
distributions in state $s$, and $n_{s,p}$ counts training perturbed cells.
The mixture and its normalization use the same set of states with available
training support. The pooled law applies to no-state settings and the
pooled-only ablation. Product coupling describes independent sampling from
the two populations, not observed cell pairs. The source-adjusted flow target is the
condition-mean response in Equation~\ref{eq:source_adjusted_target}. For a
held-out state $S$, the transfer fit mask is
$\mathcal{I}_{\mathrm{fit}}(S)=\{i:s_i\ne S\}$; clean controls from $S$ may be
used as prediction-time source context. This protocol evaluates known-target
transfer to a new cellular context.

The three frozen evaluation settings are GSE Target-LOO (19 held-out targets),
K562 Target-LOO (18 held-out targets), and GSE Cross-State-LOO (7 states by 19
targets, or 133 held-out state--target conditions). Test rows are capped at 128
cells per condition. Table~\ref{tab:protocol_objects} summarizes the prediction
controls and metric references for each protocol.
\begin{table}[!htbp]
\caption{Control pools and evaluation samples by protocol.}
\label{tab:protocol_objects}
\centering
\small
\setlength{\tabcolsep}{5pt}
\renewcommand{\arraystretch}{1.15}
\begin{tabularx}{\linewidth}{@{}>{\raggedright\arraybackslash}p{0.16\linewidth}
>{\raggedright\arraybackslash}X
>{\raggedright\arraybackslash}X
>{\raggedright\arraybackslash}p{0.18\linewidth}@{}}
\toprule
\textbf{Setting} & \textbf{Prediction controls} &
\textbf{Metric reference} & \textbf{Test rows}\\
\midrule
GSE\newline target-LOO & Cell-type grouped clean pools &
Frozen matched-control pools & Frozen target rows\\
\addlinespace[5pt]
GSE\newline cross-state-LOO & Held-out-state clean pools at prediction time &
Frozen matched-control pools & Frozen state--target rows\\
\addlinespace[5pt]
K562\newline target-LOO & Frozen global control pool &
The same frozen global pool & Frozen target rows\\
\bottomrule
\end{tabularx}
\end{table}
For GSE, the canonical evaluator forms matched-control residuals
$r^{\mathrm{match}}=x_1-\mu_{0,m}$ using frozen cell-type pools and the
observed test composition. Prediction-time controls supply model context;
test-group counts determine aggregation weights and the composition-weighted
control reference. Held-out perturbed expression is used only to compute the
observed response for scoring.

\subsection{Response residual and evaluation}
\label{app:eval_contract}
For condition $\kappa$, let $\bar x_{\kappa,g}$ and
$\overline{\widehat x}_{\kappa,g}$ denote the observed and predicted expression
means. Let $b_{\kappa,g}$ be the frozen, composition-weighted matched-control
mean in GSE, or the global DMSO mean in K562. The evaluator computes
\begin{equation}
 \Delta_{\kappa,g}=\bar x_{\kappa,g}-b_{\kappa,g},\qquad
 \widehat\Delta_{\kappa,g}=\overline{\widehat x}_{\kappa,g}-b_{\kappa,g}.
 \label{eq:app_residual}
\end{equation}
Both means use the same control reference. With the condition
index omitted, these vectors are also denoted by $\bar d$ and
$\overline{\widehat d}$ in the metric definitions below. Observed and predicted
sets may have different sizes; their means are compared without cell-to-cell
matching.

\subsection{Inference and aggregation}
\label{app:inference}
For an evaluation condition $\kappa$, let $N_{\kappa,m}$ be the supplied
number of test cells in cell-type group $m$, and define
$\mathcal{G}_\kappa=\{m:N_{\kappa,m}>0\}$. Evaluation requires a nonempty
prediction-time source pool for every group. Pool coverage is checked before
scoring, and missing groups are marked as uncovered. Each
source item is a control multiome record containing an RNA--ATAC
pair, $\bigl(x_{c,\kappa,m,i},a_{\mathrm{PROM},\kappa,m,i}\bigr)$; RNA-only
metric-reference pools are used for scoring. In the full-coverage case, let
$\mathcal{C}_{\kappa,m}=\{(x_{c,\kappa,m,i},a_{\mathrm{PROM},\kappa,m,i}):
i\in\mathcal{I}_{\kappa,m}\}$ and set
\begin{equation}
 \nu_{\kappa,m}=\frac{N_{\kappa,m}}
 {\sum_{m'}N_{\kappa,m'}}.
 \label{eq:app_weights}
\end{equation}
The benchmark supplies these observed group counts as composition metadata.
Prediction is conditional on this composition and does not include estimating
perturbation-induced changes in cell-type abundance. Alternative
aggregation rules use uniform weights
$\nu^{\mathrm{unif}}_{\kappa,m}=1/|\mathcal{G}_\kappa|$ or group frequencies
estimated from training controls.

In the GSE target-LOO condition-mean readout, the deterministic model produces
a response $\widehat\Delta_{\kappa,m,i}$ for each query control. Responses
are aggregated before adding the frozen matched-control baseline:
\begin{equation}
 \widehat\Delta_\kappa=\sum_{m\in\mathcal{G}_\kappa}\nu_{\kappa,m}
 \frac{1}{|\mathcal{I}_{\kappa,m}|}
 \sum_{i\in\mathcal{I}_{\kappa,m}}
 \widehat\Delta_{\kappa,m,i},\qquad
 \widehat\mu^{\mathrm{pert}}_\kappa=b_\kappa+\widehat\Delta_\kappa.
 \label{eq:app_point_prediction}
\end{equation}
GSE target-LOO uses matched clean groups, GSE cross-state uses held-out-state
source controls under transfer mode, and K562 uses one global source pool.
Metric-reference rows remain fixed. Appendix~\ref{app:backbone_aggregation}
compares the three response-weighting rules with the same baseline and test
cells and distinguishes prediction exports from observed cell counts.

\subsection{Metric definitions}
\label{app:metrics}
The evaluator first scores each non-empty condition and then averages over
conditions. Main comparison tables report seven metrics; ablations use
selected subsets. Auxiliary diagnostics are defined below. Higher is better
for correlations, sign accuracy, PDS, and cosine; lower is better for error
and transport distances.

\paragraph{Mean-response metrics.}
For a condition with $G$ canonical genes and matched-control mean $b$, let
$\bar x$ and $\overline{\widehat x}$ be the observed and predicted expression
means. Their mean responses are $\bar d=\bar x-b$ and
$\overline{\widehat d}=\overline{\widehat x}-b$, consistent with
Equation~\ref{eq:app_residual}. Pearson$_{\Delta}$ compares these two
vectors across genes after averaging cells within the condition. For auxiliary
sign accuracy, define $a_g=\mathbf{1}[\bar d_g>0]$ and
$\widehat a_g=\mathbf{1}[\overline{\widehat d}_g>0]$:
\begin{align}
\mathrm{Pearson}_{\Delta}&=\operatorname{Pearson}_g(\bar d,\overline{\widehat d}),\\
\mathrm{Pearson}_{\Delta,\mathrm{DE}}&=\operatorname{Pearson}_{g\in S_{20}}(\bar d_g,\overline{\widehat d}_g),
\quad S_{20}=\operatorname{Top20}_g|\bar d_g|,\\
\mathrm{acc}_{\mathrm{all}}&=G^{-1}\sum_g\mathbf{1}[a_g=\widehat a_g],\\
\mathrm{acc}_{\mathrm{de}}&=|S_{20}|^{-1}\sum_{g\in S_{20}}\mathbf{1}[a_g=\widehat a_g],\\
L2_{\mathrm{mean}}&=\lVert\bar x-\overline{\widehat x}\rVert_2,\\
\mathrm{MSE}&=G^{-1}\sum_g(\bar x_g-\overline{\widehat x}_g)^2,\qquad
\mathrm{MAE}=G^{-1}\sum_g|\bar x_g-\overline{\widehat x}_g|.
\end{align}
Pearson$_{\Delta}$ and Pearson$_{\Delta,\mathrm{DE}}$ use these
mean-response definitions throughout the paper. The DE subset is selected from the
observed mean response and is shared by all methods for that condition.
The auxiliary sign-accuracy implementation treats zero as the non-positive
sign class.

\paragraph{Auxiliary differential and identification metrics.}
For each condition, \texttt{DE\_Spearman\_Sig} first applies a Wilcoxon
rank-sum test between perturbed and matched-control cells for every gene.  The
implemented Bonferroni criterion retains genes with $p\,G<0.05$; a condition
with fewer than five retained genes contributes NaN. On the retained genes,
Spearman correlation compares the observed and predicted log-ratio scores
$\log_2((\bar x_g+\epsilon)/(b_g+\epsilon))$ and
$\log_2((\overline{\widehat x}_g+\epsilon)/(b_g+\epsilon))$.
These scores use the evaluator's normalized expression scale, with
$\epsilon>0$ for numerical stability.

For perturbation identity, \texttt{PDS} ranks the L1 distance between a
predicted residual mean and every true condition residual mean, excludes the
target gene from the distance, and converts the rank to
\begin{equation}
 \mathrm{PDS}=1-\frac{\mathrm{rank}-1}{N-1},
 \label{eq:app_pds}
\end{equation}
where $N$ is the number of conditions in the setting.

\paragraph{Transport and directional metrics.}
Observed and exported predicted residual rows are projected into one PCA space fitted on
all evaluable true residual cells in the setting (at most 50 dimensions,
random state 0).  Each condition contributes at most 64 cells to transport
calculations.  \texttt{W1\_Sinkhorn} uses
uniform weights, Euclidean PCA cost, and Sinkhorn regularization
$\mathrm{reg}=1.0$. The resulting regularized transport cost is reported as
W1 and evaluates the exported output representation specified
in Appendix~\ref{app:baselines}. \texttt{Discriminative\_Cosine} measures response direction
as the cosine similarity between the gene-space mean responses $\bar d$ and
$\overline{\widehat d}$, without PCA projection.

\subsection{Comparator implementations and prediction interfaces}
\label{app:baselines}
All comparators receive the same canonical condition identifiers and gene
adapter.  RNA and promoter-ATAC coordinates are mapped before scoring; extra
genes are dropped and missing canonical genes are handled by the method's fixed
adapter while the evaluator dimension remains fixed.

Comparators use method-specific preprocessing and optimization schedules
from their public implementations or run configurations, with adaptations
specified below. Fold-specific fits are independent. Released foundation
checkpoints remain frozen, and downstream adapters are trained on fit-fold
expression.

\paragraph{Prediction outputs and scoring.}
All point-estimation metrics use condition means. GEARS, the frozen-foundation
adapters, CPA, and the cross-state CellFlow+ATAC export provide condition means;
for W1, these means are repeated to the required number of scoring rows.
For methods that export cell samples, W1 uses those samples, while
point-estimation metrics use their means. Thus, W1 compares method-specific
output representations, whereas the primary metrics compare condition means.

\paragraph{Control.}
\textit{Control} uses unperturbed expression without a learned perturbation
adjustment. Its reported scores provide an empirical control-reference
comparison and are excluded from best/second-best learned-model highlighting.

\paragraph{GEARS.}
GEARS~\citep{roohani2024predicting} uses its published gene-relation graph and
perturbation MLP to predict post-perturbation expression. We followed the
official optimizer, preprocessing, and checkpoint rule with the published
architecture, and mapped the result to the canonical panel.

\paragraph{CellOracle and GraphVCI.}
CellOracle~\citep{kamimoto2023dissecting} uses the published
regulatory-network and vector-field simulation workflow.  GraphVCI
~\citep{wu2022predicting} uses its graph-based variational causal-inference
predictor.  We used the public implementations with their default network
construction, hidden dimensions, and regularization, and fixed graph resources
before each split.  Both methods produce condition-level gene responses after
the supplied gene mapping, with graph construction restricted to fit-fold
resources.

\paragraph{scGPT.}
scGPT~\citep{cui2024scgpt} is evaluated with its native
\texttt{TransformerGenerator} perturbation decoder, following the official
repository protocol, default preprocessing, and checkpoint rule. Generated
outputs are mapped to the canonical RNA panel before residualization.

\paragraph{scFoundation, GeneCompass, CellFM, and CellVQ.}
scFoundation~\citep{hao2024large}, GeneCompass~\citep{yang2024genecompass},
CellFM~\citep{zeng2025cellfm}, and CellVQ~\citep{wang2026illuminating} provide
frozen gene representations. A canonical GEARS-style
downstream adapter converts each representation to a condition mean, followed
by the same gene mapping and residual evaluator. We trained
the downstream adapter for 15 epochs for scFoundation (learning rate $10^{-2}$),
15 epochs for CellFM (learning rate $5\times10^{-3}$), and 5 epochs for CellVQ
(learning rate $5\times10^{-4}$). GeneCompass
uses the same frozen-embedding bridge with the optimizer settings of its public
GEARS integration.

\paragraph{CPA.}
CPA~\citep{lotfollahi2023predicting} uses its context-aware perturbation autoencoder and latent intervention
predictor, trained with the default architecture and optimization settings in
the public implementation, then decodes the predicted expression into the
canonical panel. The exported prediction is the decoded condition mean.

\paragraph{STATE.}
STATE~\citep{adduri2026predicting} uses its published latent transition model
and basal-sampled cell-set prediction interface.  We trained it from scratch
with the public implementation because the released transition checkpoints
target different dosage scenarios. The generated set is mapped to the
canonical panel. GSE cross-state-LOO
reports STATE as unavailable under the frozen export.

\paragraph{CellFlow, scDFM, PerturbDiff, and DeSCOPE.}
CellFlow~\citep{klein2025cellflow} is evaluated with the reference
flow-matching implementation, including its 50-dimensional PCA latent and ODE
integration.  scDFM~\citep{yu2026scdfm}, PerturbDiff~\citep{yuan2026perturbdiff},
and DeSCOPE~\citep{wu2026decoding} use their published inference interfaces and
released checkpoints/configurations, with adaptation restricted to the fit
fold. These methods export generated cell samples, which are mapped to the
canonical panel.

\paragraph{CellFlow+ATAC ablation.}
The ATAC-augmented CellFlow control applies train-only PCA$_{50}$ to the RNA
input and train-only SVD$_{64}$ to the gene-aligned ATAC\_PROM input, concatenates
the two representations, and passes the result to the native CellFlow
predictor. This design tests direct ATAC concatenation while retaining native
CellFlow response generation. The GSE cross-state export contains the
condition-level prediction.

\section{Method and Implementation Details}
\label{app:method_details}
\subsection{Prior feature interface and provenance}
\label{app:priors}
The prior features comprise
frozen ESM-2 protein representations~\citep{lin2023evolutionary}, frozen CLM-X
RNA-token representations~\citep{li2026clm}, TSS-centered sequence features
from GENCODE/RefTSS annotations on GRCh38
~\citep{frankish2021gencode,abugessaisa2019reftss,schneider2017evaluation}, and
locus descriptors derived from the dataset guide library, ENCODE cCRE
annotations, and lightweight JASPAR proxies
~\citep{martin2025transcription,encode2020expanded,
 castro2022jaspar}. Missing descriptors are zero-filled, and
their availability is recorded for audit. If target $p$ has $m_p>0$ guide descriptors
$\{\psi_{p,j}\}$, the pooled guide feature is
\begin{equation}
 \bar\psi_p=\frac{1}{m_p}\sum_{j=1}^{m_p}\psi_{p,j};
 \label{eq:app_guides}
\end{equation}
otherwise the guide block is zero. Target and state labels define the splits.
Control ATAC state summaries use a
control-only TF--IDF/LSI
transform~\citep{stuart2021single}. Any control-derived accessibility summary
or response-normalization quantity is recomputed within the active training fold,
including in held-out-state transfer. Only frozen embeddings and static
annotations specified before splitting are shared across folds.

The target-locus condition is
\begin{equation}
 \mathbf p_i=\operatorname{concat}
 \left(e^{\mathrm{ESM2}}_p,e^{\mathrm{CLMX}}_p,
 f^{\mathrm{seq}}_p,f^{\mathrm{reg}}_p\right),
 \label{eq:app_prior_matrix}
\end{equation}
where the two target embeddings, 96-dimensional sequence summary, and
128-dimensional regulatory summary enter the target block in
Equation~\ref{eq:perturb_condition}. The stored 265-coordinate regulatory
descriptor includes repeated summary and composite coordinates and is reduced
as a single feature bank. Static locus annotations and
fold-derived control summaries remain separate, and the input hashes and
projection scope are recorded in the release audit. The resulting pooled and
state-aware flow conditions have 576 and 736 dimensions, respectively.

Each external-prior ablation in Table~\ref{tab:ablation_priors} replaces the named
ESM-2, CLM-X, TSS-sequence, or regulatory block with zeros when constructing
$c_i$, retaining the same network capacity.

The CGRR source bank is rebuilt for each fold. Its projection map, ATAC prototypes
$P_s$, control means $\mu_s^0$, and control-state offsets $B_s$ are rebuilt from
training-fold controls. In Target-LOO, target/locus features retrieve permitted
training perturbations whose observed compositions define the query mixture.
In Cross-State, the clean query ATAC selects among the fixed prototypes.
Equation~\ref{eq:cgrr_source} combines the control-state offsets using the weights
returned by the route for that protocol.
If a protocol specifies a fixed support or feature-distance gate, the gate is
applied before normalizing the routing weights; the source bank is unchanged.
Training mixtures use the observed perturbation composition across source
groups, with control frequencies as a fallback when support is insufficient.
These observed weights define the source-adjusted training endpoint. At
inference, Equation~\ref{eq:cgrr_source} instead uses query-dependent weights
from target/locus similarity for Target-LOO or clean-control ATAC similarity
for Cross-State. Training and inference therefore use different weighting rules.
Cross-State reuses the ATAC weights to combine state-centered responses
$A_{p,s}$; it does not add the routed source offset to the final prediction
(Equation~\ref{eq:response_composition}).

\subsection{Flow target and numerical details}
\label{app:flow}
For the pooled source-on candidate, CGRR first
constructs $d_i^{\mathrm{source}}$ using training-fold controls. Writing
$T_\phi$ for the complete PCA-anchored transport operator, the flow target is
the source-adjusted response and its decoded displacement is added back to the
routed source:
\begin{equation}
 \widetilde\Delta_i=\Delta_i-d_i^{\mathrm{source}},
 \qquad a_i=T_\phi(\widetilde\Delta_i\mid c_i),
 \qquad \Delta_i^{\mathrm{base}}=d_i^{\mathrm{source}}+a_i.
 \label{eq:app_flow_target}
\end{equation}
The flow uses the training-control mean as the PCA anchor:
\begin{equation}
 z_0=E_{\mathrm{PCA}}(\bar b_{\mathrm{tr}}),
 \qquad z_{1,i}=E_{\mathrm{PCA}}(\bar b_{\mathrm{tr}}+\widetilde\Delta_i).
 \label{eq:app_flow_endpoints}
\end{equation}
The linear conditional-flow path is
\begin{equation}
 z_{t,i}=z_0+t(z_{1,i}-z_0),
 \qquad u_i^*=z_{1,i}-z_0,
 \qquad t\sim\mathcal{U}(0,1),
 \label{eq:app_flow_path}
\end{equation}
and optimization minimizes the squared velocity error in
Equation~\ref{eq:delta_path}. The velocity field is conditioned on the pooled
or state-aware target--locus condition. CGRR uses target/locus or clean-control
ATAC queries according to the protocol. At test time, a fixed-step solver
integrates from the deterministic control-RNA anchor. For $K$ Euler steps,
$h=1/K$ and
\begin{equation}
 z^{(k+1)}=z^{(k)}+h\,v_\phi(z^{(k)},kh,c_i),
 \qquad z^{(0)}=z_0,\quad k=0,\ldots,K-1,
 \label{eq:app_euler}
\end{equation}
Each step uses one velocity evaluation, so $K$ is the NFE, with default
$K=8$. The decoded displacement and stabilized source-on candidate are
\begin{equation}
 a_i=D_{\mathrm{PCA}}(z^{(K)})-D_{\mathrm{PCA}}(z_0),
 \qquad
 \widehat\Delta_i=\mathcal C(\Delta_i^{\mathrm{base}}).
 \label{eq:app_ode}
\end{equation}
The transport flow applies the configured mean blend $\rho=0.18$ to its decoded
displacement; the state-expanded branch uses state-centered rows and the same
blend. The reliability gate is fitted from training rows and evaluated without
held-out perturbation responses.
Writing the affine PCA reconstruction as
\begin{equation}
 D_{\mathrm{PCA}}(z)=\mu_{\mathrm{PCA}}+Bz,
 \qquad B\in\mathbb{R}^{G\times128},
 \qquad a_i=B\bigl(z^{(K)}-z_0\bigr),
 \label{eq:app_pca_displacement}
\end{equation}
shows that the PCA offset cancels in the decoded displacement. The candidate
preserving operator then acts on the combined source and flow base.
Let $\mathcal M_{\mathrm{tr}}$ be the union of candidate genes and genes
protected by training-fold DE support, and let
$\bar\Delta_{\mathrm{tr}}$ be the mean training response. The canonical
candidate-preserving operator is
\begin{equation}
  \widehat\Delta_{i,g}=\mathcal C(\Delta_i^{\mathrm{base}})_g=
 \begin{cases}
   \Delta_{i,g}^{\mathrm{base}}, & g\in\mathcal M_{\mathrm{tr}},\\
   \alpha \Delta_{i,g}^{\mathrm{base}}+(1-\alpha)\bar\Delta_{\mathrm{tr},g},
  & g\notin\mathcal M_{\mathrm{tr}},
 \end{cases}
 \qquad \alpha=0.10.
 \label{eq:candidate_stabilization}
\end{equation}
The operator uses no held-out response values. This source-on candidate is
one input to the Target-LOO reliability gate. The state-centered branch
instead uses $r_{i,s}$ from Equation~\ref{eq:state_response} as its flow target,
without source addback. Protocol-specific reconstruction follows
Equation~\ref{eq:response_composition}, adding the matched-control baseline
once after candidate selection or ATAC-weighted state transfer.
The velocity-network architecture and default training configuration are
specified in Appendix~\ref{app:implementation} and
Table~\ref{tab:app_default_hyperparameters}.

\subsection{Canonical implementation and training configuration}
\label{app:implementation}

\paragraph{Configuration and fold construction.}
Table~\ref{tab:app_default_hyperparameters} gives the primary ChromaPert
configuration for the GSE protocols. Each fold holds out one target or state
and fits the model using the remaining targets and permitted controls. PCA is
fitted once on outer-training control RNA and then fixed. Target embeddings are frozen;
condition normalization, source summaries, and CGRR routes are fold-specific.
Preprocessing and controls follow Appendices~\ref{app:datasets}--\ref{app:protocol}.

\paragraph{Evaluation-setting instantiation.}
ChromaPert uses the same pooled-reference decomposition and CGRR source bank in
every setting. A conditional transport flow is fit on source-adjusted deltas, a
state-expanded transport branch is available when state summaries are used, and
Target-LOO applies a train-only reliability gate to the decoded candidate bases;
Cross-State uses the statebase branch directly.
Unseen-target settings use the target/locus context
route to infer a training-state mixture. Held-out-state transfer uses
clean-control ATAC to match the query to the training prototypes. When
population composition is supplied, state-specific predictions are aggregated
with those counts; state-transfer evaluation returns the selected state
prediction.

\paragraph{Velocity network and optimization.}
The velocity MLP concatenates $z_t$, $t$, and the pooled or state-aware
condition $c$ (576 or 736 dimensions, respectively).
LayerNorm precedes two width-288 hidden linear layers, each with GELU and
dropout, followed by a linear 128-dimensional velocity output. AdamW minimizes
Equation~\ref{eq:delta_path} for the setting-specific epoch budgets at a
constant learning rate.
Direction/norm and supervised ATAC loss weights are zero. Only the transport
velocity MLPs are trained; PCA and the source bank remain fixed.

\paragraph{Primary settings.}
Table~\ref{tab:app_default_hyperparameters} specifies the primary
configuration. We report temperature and blend sensitivity
(Tables~\ref{tab:ablation_temperature} and~\ref{tab:ablation_blend}) and
three-seed variation (Appendix~\ref{app:multiseed}) separately.
These analyses characterize performance under the tested configurations and
the selected primary setting.

\section{Discussion and Limitations}
\label{sec:discussion}

ChromaPert connects perturbation targets to cellular regulatory context
through chromatin-aware conditioning and source-routed conditional transport
flow modeling.
Across the three evaluation settings, it achieves the highest all-gene
response correlation across the benchmark methods. Ablations
show the complementary value of molecular target priors and regulatory
context, while biological analyses show that predictions retain
chromatin-associated and state-resolved response structure.

The main limitation is the requirement for paired scRNA-seq and scATAC-seq
control profiles with a compatible promoter-accessibility mapping,
restricting direct application when matched multimodal controls are
unavailable. The framework predicts post-perturbation expression, with population
readouts conditional on supplied state proportions.
Future work will extend ChromaPert to unpaired or single-modality inputs
and incorporate distal regulatory information.

\begin{table}[!htbp]
\caption{Default configuration for ChromaPert on GSE target-LOO.}
\label{tab:app_default_hyperparameters}
\centering
\footnotesize
\setlength{\tabcolsep}{6pt}
\renewcommand{\arraystretch}{1.03}
\begin{tabularx}{\linewidth}{@{}>{\raggedright\arraybackslash}X
>{\raggedright\arraybackslash}p{0.40\linewidth}@{}}
\toprule
\textbf{Parameter} & \textbf{Default} \\
\midrule
\multicolumn{2}{@{}l}{\textbf{Data and representations}} \\
Held-out targets / canonical RNA genes & 19 / 2,013 \\
Random seed & 114 \\
ESM-2 / CLM-X dimensions & 64 / 64 \\
TSS-sequence / regulatory dimensions & 96 / 128 \\
Target--locus feature dimension & 352 \\
Static locus block dimension & 224 \\
Control RNA / ATAC state dimensions & 96 / 64 \\
Pooled / state-aware flow condition & 576 / 736 \\
PCA latent / velocity-output dimensions & 128 / 128 \\
\midrule
\multicolumn{2}{@{}l}{\textbf{Velocity architecture and optimization}} \\
Hidden layers / width & 2 / 288 \\
Input normalization / activation & LayerNorm / GELU \\
Dropout probability (flow) & 0.11 \\
Optimizer & AdamW \\
Training epochs (flow) & 360 \\
Batch size & 64 \\
Learning rate & $8\times10^{-4}$ \\
Weight decay & $10^{-4}$ \\
Learning-rate schedule & Constant (no scheduler) \\
\midrule
\multicolumn{2}{@{}l}{\textbf{Routing, integration, and evaluation}} \\
CGRR feature temperature / ATAC temperature & 0.20 / 0.15 \\
ATAC query parameter for Cross-State (\texttt{atac\_top\_k}) & 50 \\
Transport-mean blend & 0.18 \\
Candidate-preserving prior $\alpha$ & 0.10 \\
ODE solver / function evaluations & Euler / 8 \\
Condition-response aggregation & Official/test composition \\
Genes for Pearson$_{\Delta,\mathrm{DE}}$ & Top 20 absolute mean responses \\
\bottomrule
\end{tabularx}
\end{table}

\end{document}